\documentclass[10pt,journal,compsoc]{IEEEtran}

\ifCLASSOPTIONcompsoc
  \usepackage[nocompress]{cite}
\else
  \usepackage{cite}
\fi
\ifCLASSINFOpdf
  \usepackage[pdftex]{graphicx}
  \DeclareGraphicsExtensions{.pdf,.jpeg,.png,.jpg}
\else
  \usepackage[dvips]{graphicx}

\fi
\ifCLASSOPTIONcompsoc
  \usepackage[caption=false,font=footnotesize,labelfont=sf,textfont=sf]{subfig}
\else
  \usepackage[caption=false,font=footnotesize]{subfig}
\fi

\newcommand{\RQone}{How effective is \fs at validating  true vulnerabilities across slice levels?}
\newcommand{\RQtwo}{How effective is \fs at filtering out false alarms across slice levels?}
\newcommand{\RQthree}{How accurately does \fs reproduce the actual execution paths of true vulnerabilities in stack traces?}
\newcommand{\RQfour}{How efficient is \fs with its slice-level optimizations?}
\newcommand{\RQfive}{How does \fs compare with directed fuzzers in efficiency and target reachability?}
\newcommand{\RQsix}{How does \fs compare with LLM based approaches for classifying vulnerabilities?}
\newcommand{\RQseven}{How does \fs compare with slice based approach \oldfs?}

\usepackage{pbalance}

\usepackage{booktabs} 
\usepackage{pbalance}
\usepackage{orcidlink}
\usepackage{caption}
\usepackage{subcaption}
\usepackage{multirow}
\usepackage{tabularx}
\usepackage{url}
\usepackage{balance}
\usepackage[font=small]{caption}
\usepackage{comment}
\usepackage{makecell}

\usepackage{comment}
\usepackage{makecell}

\usepackage{tikz}
\usepackage{pgfplots}
\pgfplotsset{compat=1.18}
\usepgfplotslibrary{statistics}
\usepackage{enumitem}
\usepackage{fancyhdr}
\usepackage{algorithm}
\usepackage{algpseudocode}
\usepackage{threeparttable}
\usepackage{xspace}
\usepackage{soul}
\usepackage{color,xcolor}
\usepackage{listings}
\usepackage[most]{tcolorbox}
\usepackage{hyperref}

\newcommand{\oldfs}{\textsc{FuzzSlice}\xspace}
\newcommand{\fs}{{\textsc{{SnipTest}}}\xspace}

\tcbset{
  stepbox/.style={
    colback=gray!20,
    colframe=gray,
    boxrule=0pt,
    arc=0mm,
    left=1mm, right=1mm, top=0.5mm, bottom=0.5mm,
    fontupper=\bfseries,
    enhanced,
    sharp corners,
  },
  myrqstyle/.style={
    enhanced,
    colback=gray!10,
    colframe=gray!30,
    boxrule=0.4pt,
    arc=1mm,
    left=4pt,
    right=4pt,
    top=2pt,
    bottom=2pt,
    boxsep=4pt,
    fontupper=\bfseries\scshape,
    before skip=6pt,
    after skip=4pt,
  }
}

\lstdefinestyle{mystyle}{
  backgroundcolor=\color{backcolour},
  commentstyle=\color{codegreen},
  keywordstyle=\color{magenta},
  numberstyle=\tiny\color{codegray},
  stringstyle=\color{codepurple},
  basicstyle=\ttfamily\footnotesize,
  breakatwhitespace=false,
  breaklines=false,
  captionpos=b,
  keepspaces=true,
  numbers=left,
  numbersep=5pt,
  showspaces=false,
  showstringspaces=false,
  showtabs=false,
  tabsize=2,
  frame=single,
  rulecolor=\color{bordercolour},
  framerule=0.8pt,
  columns=fullflexible,
  xleftmargin=1.5em,
  framexleftmargin=1.5em
}

\lstdefinestyle{mystyle}{
  backgroundcolor=\color{backcolour},
  commentstyle=\color{codegreen},
  keywordstyle=\color{magenta},
  numberstyle=\tiny\color{codegray},
  stringstyle=\color{codepurple},
  basicstyle=\ttfamily\footnotesize,
  breakatwhitespace=false,
  breaklines=false,
  captionpos=b,
  keepspaces=true,
  numbers=left,
  numbersep=5pt,
  showspaces=false,
  showstringspaces=false,
  showtabs=false,
  tabsize=2,
  frame=single,
  rulecolor=\color{bordercolour},
  framerule=0.8pt,
  columns=fullflexible,
  xleftmargin=1.5em,
  framexleftmargin=1.5em
}

\newtcolorbox{boldheading}{
  colback=white, colframe=black, fontupper=\large, boxrule=1pt
}

\definecolor{codegreen}{rgb}{0,0.6,0}
\definecolor{codegray}{rgb}{0.5,0.5,0.5}
\definecolor{codepurple}{rgb}{0.58,0,0.82}
\definecolor{backcolour}{rgb}{1,1,1} 
\definecolor{bordercolour}{rgb}{0.8,0.8,0.8} 

\begin{document}
%
\title{\fs: Fuzzing Multi-Level Code Slices for Validating Vulnerabilities}

\IEEEtitleabstractindextext{%
\begin{abstract}
Modern software systems are increasingly complex, and static analysis tools are commonly used to identify potentially vulnerable code by issuing warnings.
However, these warnings often require manual inspection to confirm whether the reported issues are real, making the process time-consuming and error-prone.
Directed fuzzing has emerged as a powerful automated technique to validate the warnings.
However, applying it to the entire project in response to each warning is computationally infeasible, often requiring days of execution to achieve only incremental improvements in code coverage.

We present \fs, an execution-based warning triage framework that generates and fuzzes compiled code slices centered around static-analysis warnings. 
Rather than proving exploitability in the full program, \fs provides evidence about how a warning behaves under progressively expanded sliced execution contexts.
Unlike prior approaches that extract slices from the program entry point or limit slice size for filtering, \fs constructs slices of arbitrary size and compiles them into standalone testable units. 
It employs a {layer-by-layer slicing strategy}, incrementally expanding context around the target location to validate potential vulnerabilities with increasing precision.
We evaluate \fs on a benchmark of 97 true vulnerabilities and 97 false alarms across three real-world projects. 
\textcolor{black}{\fs produces Possible True Positive evidence for 53 of 97 confirmed vulnerabilities (54.6\%) by triggering the corresponding bug oracle consistently across all three analyzed slice levels, while the remaining cases are unreachable.}
Particularly, in 40.2\% of these cases, it exploits the vulnerability along the observed execution path, matching the top three stack frames. 
\textcolor{black}{On the 97 confirmed false alarms, \fs produces Possible False Positive evidence for 54 cases (55.6\%) by reaching the warning without triggering the bug oracle, but misclassifies 28 cases (28.8\%),and the remaining cases are unreached.}
\textcolor{black}{Compared with AFLGo under our benchmark configuration, unseeded \fs reaches more locations than unseeded AFLGo and achieves comparable reachability to seeded AFLGo.}
Furthermore, \fs enhances efficiency, achieving a $5.5$--$10.6\times$ speedup in fuzzing time compared to seeded or unseeded directed fuzzers.
We also compare \fs with LLM based approach for classifying static analysis warnings LLM4SA. We find that \fs outperforms LLM4SA with an F1 score of 0.791 compared to 0.360 for LLM4SA.
Finally, we demonstrate the practical relevance of \fs by identifying three new vulnerabilities in two open source projects leading to disclosure of CVE-2025-11964.
\end{abstract}

\begin{IEEEkeywords}
Fuzzing, Vulnerability detection, Code slicing
\end{IEEEkeywords}}

\author{
    \IEEEauthorblockN{Aniruddhan Murali\orcidlink{0000-0002-4405-1657}, \IEEEmembership{Student Member, IEEE}, Noble Saji Mathews\orcidlink{0000-0003-2266-8848}, \IEEEmembership{Student Member, IEEE},
    \\Mahmoud Alfadel\orcidlink{0000-0002-2621-6104}, \IEEEmembership{Member, IEEE}, Meng Xu\orcidlink{0009-0001-6364-4837} \IEEEmembership{Member, IEEE} and Meiyappan Nagappan\orcidlink{0000-0003-4533-4728}~\thanks{Mahmoud Alfadel is with the University of Calgary, Canada. The remaining authors are with the David R. Cheriton School of Computer Science, University of Waterloo, Canada. 
    \\E-mail:\{a25murali, noblesaji.mathews, mei.nagappan, meng.xu.cs\}@uwaterloo.ca, mahmoud.alfadel@ucalgary.ca}}
}

\maketitle

\IEEEdisplaynontitleabstractindextext

%
\IEEEpeerreviewmaketitle

\section{Introduction}
\label{sec:introduction}

Static analysis tools are widely used to detect potentially harmful bugs (i.e., vulnerabilities) early in the software development lifecycle~\cite{Ayewah2008Aug, louridas2006static, zheng2006value, vassallo2020developers}.
However, they often generate a large volume of warnings, many of which are false positives~\cite{Kang2022May,Nadeem2012Mar,Park2016May,Aloraini2017Sep}.
In parallel, new code are frequently committed into the codebase, and these changes must be carefully reviewed to prevent regressions in production systems~\cite{eyolfson2014correlations}.
As a result, developers face a costly and error-prone triage process to determine which warnings or commits require urgent attention and which can be deprioritized.

A promising idea is dynamic validation---a family of techniques that attempt to confirm whether a suspicious code location actually results in failures at runtime. 
Among these, fuzzing is particularly intriguing due to its effectiveness in uncovering vulnerabilities~\cite{Li2018Dec, csallner2005check}. However, applying fuzzing to large codebases or individual warnings is challenging:
writing tailored harnesses for each warning is labor-intensive; while general fuzzing often reach a coverage plateau before reaching the intended code location~\cite{ bohme2020fuzzing}. 

These limitations have led to the emergence of directed fuzzing, which is further categorized into distance-based~\cite{bohme2017directed, Chen2018Oct, DAFL} and slicing-based.
In slicing-based fuzzing, a program slice (i.e., code closely related to the warning or commit) is isolated and fuzzed independently~\cite{FuzzSlice, KallingalJoshy2021Jul, harman1995using, Chen2022Nov}.
Compared with distance-based fuzzing, slicing-based fuzzing starts at code locations that are much closer to the suspicious code, which significantly increases the chances of hitting the target.
While one cannot get an end-to-end input that proves the existence of a bug, the under-constrained nature of a slice is especially useful for ruling out false bug reports, i.e., if a bug cannot be triggered with even less constraints on the inputs, it is more likely to be a false alarm.
This is the core idea behind \oldfs~\cite{FuzzSlice}.
%

However, current slicing-based techniques face a key limitation.
Program slices are generated ahead and independent of fuzzing.
More specifically, the scope of slices is often predetermined based on heuristics and does not adapt to fuzzing results.
This implies that the fuzzer may either omit necessary context (if the slice is too small~\cite{FuzzSlice}) or introduce excessive complexity (if the slice is too large~\cite{bohme2017directed}). 
Furthermore, predetermined slices imply a limited set of execution paths or entry points, which can lead to false negatives by ignoring alternative call chains that may reach the same vulnerability ~\cite{garg2018empirical}.
These constraints can limit their ability to expose exploitable vulnerabilities or correctly dismiss benign warnings.

To address these challenges, we introduce \fs, a compositional fuzzing framework that validates vulnerabilities through slices of increasing context.
Instead of relying on a fixed set of slices, \fs organizes validation around \emph{slice levels}. At Level~0, \fs generates a slice rooted at the function that contains the warning. At Level~1, it expands the analysis to immediate callers of that function and their dependencies. At Level~2, it further expands to higher-level callers, and so on.
\textcolor{black}{Critically, this leveled design is not merely about testing larger slices. The power lies in treating the trajectory of outcomes across levels as a diagnostic signal. A warning that triggers only in the most underconstrained slice (Level 0) is likely brittle and context‑sensitive—it may disappear once caller guards or initialization operations are reintroduced. In contrast, a warning that persists as we expand the calling context points to a systemic issue that warrants prioritization. By leveraging the sequence of outcomes rather than any single slice in isolation, \fs elevates slice‑based fuzzing from a filtering technique to reactive slicing for distinguishing benign anomalies from genuinely critical vulnerabilities.}
Figure~\ref{img:levels} illustrates this process using a call graph in which the vulnerable function is progressively embedded in broader calling contexts.

Our work \fs improves on \oldfs~\cite{FuzzSlice} to support multi-level slicing—the idea of progressively expanding slices backward from a vulnerability.
Multi-level slicing introduces new challenges such as handling an explosion of paths at higher levels, and avoiding redundant fuzzing across similar slices.
\fs tackles these challenges through dedicated optimizations, time-bounded slicing, and a staged evaluation design that makes multi-level analysis tractable and meaningful (Sections~\ref{sec:approach} \& ~\ref{sec:optimizations}).
\fs tests each bug across multiple slice levels to assess not just whether it can trigger a vulnerability, but how consistently it does so across all slices, providing a clearer indication of the bug likelihood to represent an exploitable vulnerability.

\textcolor{black}{Throughout the paper, we use “validation” in a pragmatic sense, i.e., \fs produces execution evidence for warning triage in slice context, rather than proving full-program exploitability. A warning classified as a possible true positive indicates that \fs triggered the corresponding bug oracle within the analyzed sliced contexts, while a possible false positive indicates that the warning was reached without triggering the bug oracle in those contexts.}

Our study contributes the following advances:
\begin{itemize}
\item \textbf{Multi-context warning triage.} \textcolor{black}{We introduce a warning-triage protocol that organizes executable slices by call-graph distance from a warning. \fs changes the unit of reasoning from one local function slice to a family of progressively constrained execution contexts. It aggregates observations within each level and across levels to determine whether the warning persists, disappears, or becomes unreachable as caller context is restored. To the best of our knowledge, \fs is the first work to dynamically adjust code slices based on fuzzing feedback.}
\item \textbf{Engineering support for scalable multi-level analysis.} \textcolor{black}{We introduce caching, pruning, staged stopping, parallel execution and wrapper-variable trimming to make the construction and fuzzing of multiple caller-rooted slices practical. These are engineering extensions around the low-level slice construction inherited from \oldfs.}
\item \textbf{Comprehensive evaluation.} We evaluate \fs across four key dimensions: effectiveness, fidelity, efficiency in detecting true vulnerabilities and false alarms. Our work is comprehensive due to the indepth discussion of true positives, true negatives, false positives and false negatives - rigor not commonly found in prior work~\cite{FuzzSlice, TPSlice, ramos2015under}. We also provide direct comparison with the directed fuzzer AFLGo~\cite{bohme2017directed} and the  LLM-based approach LLM4SA~\cite{LLM4SA}. \fs achieves a 5.5-10.6x speedup in fuzzing time over seeded or unseeded AFL-Go. Additionally, \fs outperforms LLM4SA, with an F1 score of 0.791 versus 0.360.
\item \textbf{Practical vulnerability detection.} We demonstrate the practicality of \fs by detecting 3 real world vulnerabilities in 2 Open Source projects~\cite{vim-buffer, vim-null, libpcap-bug} leading to disclosure of CVE-2025-11964~\cite{CVE}.
\item \textbf{Artifact \& Reproducibility.} Our replication package and datasets are publicly \href{https://osf.io/cjrtg/?view_only=da5efa6b27d0484193eb6c9343324bf0}{available}.
\end{itemize}

\section{Background on Fuzzing Approaches}
In this section, we provide background on coverage-guided fuzzers and directed fuzzers.

\textbf{Coverage-guided fuzzers.} Fuzzing is an automated software testing technique that generates program inputs to expose crashes, bugs, and security vulnerabilities. Traditional fuzzers operate in an undirected manner: they aim to maximize overall code coverage or execution diversity without prioritizing any specific program locations. Popular coverage-guided fuzzers, such as AFL and libFuzzer~\cite{BibEntry2023Mar, BibEntry2023Jan}, fall into this category and have been widely successful in discovering shallow and medium-depth bugs.

\textbf{Directed fuzzers.} Directed fuzzing extends this model by biasing input generation toward specific targets, such as known vulnerable code locations, patches, or warnings reported by static analysis tools. Instead of maximizing global coverage, directed fuzzers such as AFLGo and HawkEye \cite{bohme2017directed, Huang2022May}, guide execution toward target locations using metrics such as control-flow distance or reachability. This approach is particularly effective for deep or hard-to-reach bugs but typically requires accurate target specification and may incur additional overhead in analysis and guidance.

\section{Why \fs? A Motivating Example}
\label{sec:motivation}

Our goal is to validate whether suspicious code locations, such as those emerged by static analysis tools or recently changed, correspond to actual vulnerabilities.
For instance, if a tool reports a potential memory leak or buffer overflow, we aim to determine whether the issue is a true bug or a false alarm. 
Manually inspecting each warning or fuzzing the entire program in response to each one is inefficient and often ineffective~\cite{Bohme2020Nov, bohme2020fuzzing}.

\begin{figure}[tb!]
    \centering {\includegraphics[width=\columnwidth]{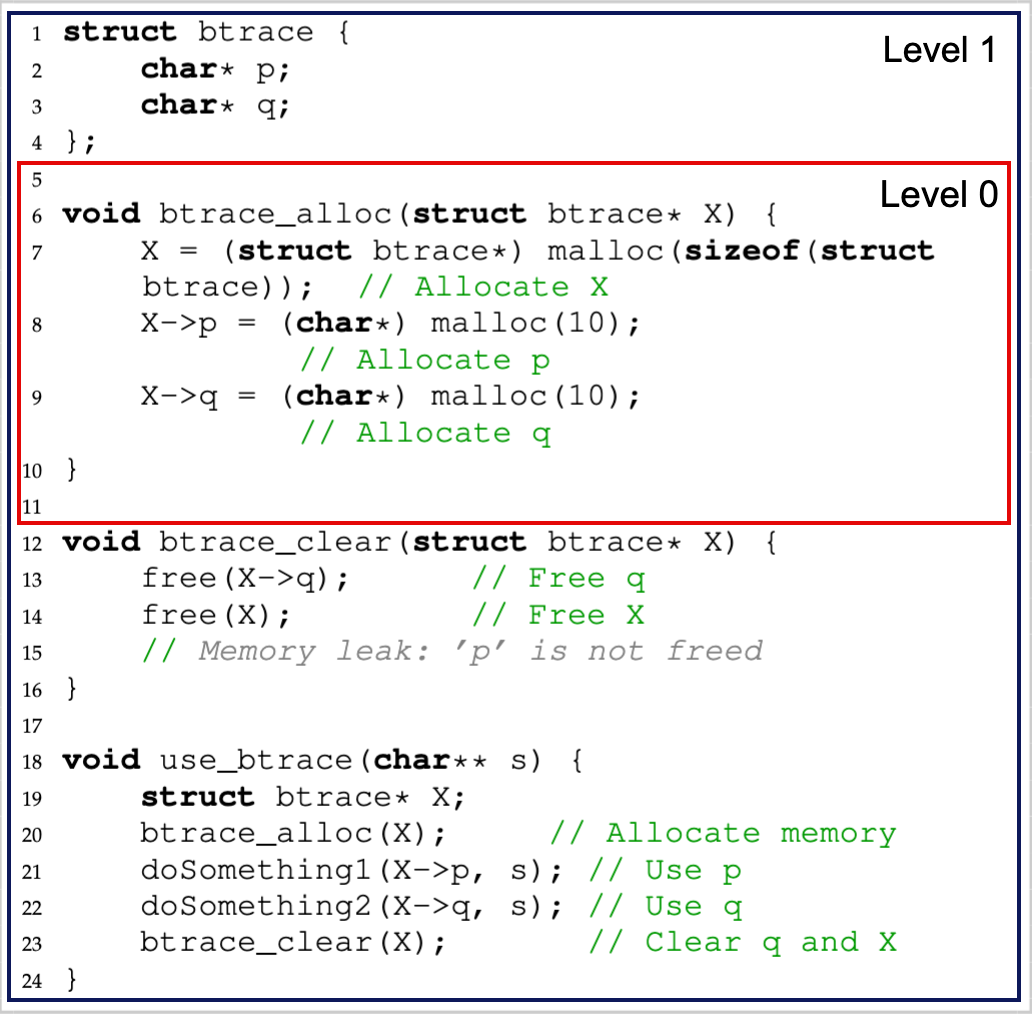}}   
    \caption{Example from \texttt{binutils} project~\cite{example-leak}. The pointer \texttt{p} is allocated but never freed, leading to a memory leak.}
    \label{lst:motivating-example}
\end{figure}

To illustrate our approach, we present a motivating example from the \texttt{binutils} project~\cite{example-leak}. Listing~\ref{lst:motivating-example} shows simplified C code involving three functions: the function \verb|btrace_alloc| performs heap allocation for a struct and its fields, \verb|btrace_clear| partially deallocates the structure, and \verb|use_btrace| orchestrates the full flow.





When a static analysis tool such as Infer~\cite{infer} analyzes this code, it reports a memory leak associated with the field \verb|p| in the \verb|btrace_alloc| function. 
Validating this warning using whole-program fuzzing, such as fuzzing the entire \texttt{binutils} project, is slow, often requiring days of execution and project-specific fuzzing dictionaries, and may still fail to trigger the vulnerability.

\fs adopts a more targeted strategy.
It begins by generating and fuzzing a minimal code slice containing the warning location, which we refer to as {Level~0}. 
In this example, Level~0 includes only the function \verb|btrace_alloc|. 
When fuzzed in isolation, a bug oracle e.g. LeakSanitizer~\cite{leaksanitizer}  reports memory leaks for all three heap-allocated components: \verb|p|, \verb|q|, and \verb|X|.

To improve diagnostic precision, \fs incrementally expands the slice.
At {Level~1}, \fs includes the caller \verb|use_btrace| and its dependencies, such as \verb|btrace_clear|. 
This broader slice enables both allocation and deallocation logic to be exercised.
Fuzzing at this level suppresses the leak reports for \verb|q| and \verb|X|, as valid deallocation paths are now included. 
However, the warning for \verb|p| persists, since \verb|p| is never freed. At higher levels, we would still expect \verb|p| to be leaked.

This example highlights the core insight behind \fs: by observing how warnings behave across incrementally expanded slice levels, we can distinguish between true bugs and false alarms. 
If a warning consistently triggers across multiple levels, it is likely signaling a true vulnerability.
If the code is executed but the warning does not manifest, it is likely a false alarm.
That is, iteratively growing the code slice around the warning and checking if the warning persists, \fs can offer a lightweight yet semantically grounded way to prioritize warnings without requiring full-program fuzzing.

\section{Related Work}
Program slicing is well-studied, with most work focusing on uncompiled backward slices from entry points to vulnerabilities~\cite{lei2013backward, binkley1996program, korel1988dynamic}. However, these slices are often overly large for arbitrary code points~\cite{Binkley1, sridharan2007thin, Binkley, jaffar2012path}. In contrast, \fs operates at the function level, terminating backward slices at different functions, and finally compiling and linking small slices to enable quick exploration.

A common technique for dynamically validating vulnerabilities is directed fuzzing.
Prior work has proposed various strategies for directing fuzzers toward specific program regions~\cite{272157,Huang2022May,Wustholz2019May,Chen2018Oct,Bohme2017Oct,Razzer}, primarily by mutating inputs to increase the likelihood of reaching a target location.
Therefore directed fuzzers can be used to verify static analysis warnings.
The key distinction between \fs and directed fuzzing lies in their entry points: while directed fuzzers typically begin at the \textsc{main} function and explore the entire program, \fs generates small slices around warning locations and confines fuzzing within these slices.  
Input mutation strategies from directed fuzzing are orthogonal to \fs and could be integrated into its fuzzing engine.



\textcolor{black}{\oldfs is the closest implementation and methodological predecessor to \fs. 
Given a target function, \oldfs resolves dependencies, extracts a compilable function-rooted slice, generates a fuzzing wrapper, and tests the resulting binary. 
Its unit of reasoning is a single underconstrained local slice, and its primary use case is false-positive filtering. \fs reuses this slice-materialization operation—it does not introduce a new low-level slicing algorithm. 
Rather, its contribution lies in the surrounding framework: selecting multiple caller roots via static call-graph analysis, grouping slices by caller distance, testing multiple contexts per level, aggregating observations within and across levels, and applying staged stopping and scalability optimizations. 
Level 0 is the closest comparable setting to \oldfs, as both analyze a slice rooted at the warning-containing function. However, \fs is not identical even at Level 0, owing to wrapper-generation improvements. 
More importantly, Level 0 is only the initial step in a multi-level warning-triage process. While \oldfs focuses on underconstrained exploration of fine-grained slices primarily for false-positive filtering, \fs supports adjustable slice sizes, enabling improved precision and more effective vulnerability detection. 
Beyond that, \fs incorporates static analysis to automatically discover ancestor functions, expands slices based on fuzzing outcomes, and introduces multiple optimizations to classify warning behavior across progressively larger code contexts as benign, vulnerable, or unreachable.}

Runtime bug oracles such as AddressSanitizer, LeakSanitizer, UndefinedBehaviourSanitizer and MemorySanitizer~\cite{serebryany2012addresssanitizer, LSAN, UBSAN, MSAN} are used by \fs to detect several issues (e.g., out-of-bounds access, null pointer dereference, timeouts, double free), but other oracles—such as developer-provided assertions or even LLM-based assertions—can be integrated at runtime without modifying the framework. The design of effective oracles is complementary to, and independent of, \fs itself. As more expressive sanitizers or assertions become available, \fs (like any fuzzing system) can naturally detect a broader class of bugs, including logic errors.

Other approaches for identifying actionable static-analysis warnings include machine-learning–based techniques and, more recently, LLM-based approaches. Machine-learning–based methods train classifiers to distinguish vulnerable from non-vulnerable code using features extracted from static-analysis outputs, such as warning types, code metrics, or data-flow characteristics, and they require large, high-quality labeled datasets~\cite{hanam2014finding,yang2021learning, yedida2023find}. LLM-based approaches, in contrast, leverage pretrained language models to reason about code semantics and warning context, reducing the need for handcrafted features but still relying heavily on the quality and availability of static-analysis information or prompt context~\cite{wen2024automatically, li2024enhancing}. These approaches differ fundamentally from \fs; they operate at the prediction level without executing the program.

\section{\fs Approach}
\label{sec:approach}

\fs aims to validate potential vulnerabilities reported by static analysis tools or recently changed code by generating and testing executable code slices that include the reported vulnerability across various program contexts.
This is done through four key steps:  
\textbf{(1)} Construct a static function-level call graph;  
\textbf{(2)} Identify the function enclosing the vulnerability;  
\textbf{(3)} Generate and compile slices of increasing context;  
\textbf{(4)} Fuzz the resulting binaries and classify the vulnerability.  
We describe each step below.

\noindent
\textbf{{(1) Construct a static function-level call graph.}} To assess whether a vulnerability is likely to be triggered at runtime, we first need to understand the possible execution paths in the program. 
Therefore, we build a static call graph that maps caller-callee relationships between functions.
We use Tree-sitter~\cite{treesitter}, to extract the Abstract Syntax Tree (AST), identifying function definitions and calls to build the function-level call graph.

\vspace{1mm}
\noindent
\textbf{{(2) Identify the function enclosing the vulnerability.}}
Static analysis tools often generate warnings at specific code locations, but without runtime context.
Hence, identifying the function enclosing the  vulnerability is necessary to isolate the affected code region.
In this step, using the AST from Step 1,
we search for the exact function enclosing the given warning and designate it as the vulnerable function.
This function forms the starting point for all code slices.

\vspace{1mm}
 \noindent
\textbf{{(3) Generate and compile slices of increasing context.}}
Analyzing a vulnerability warning in isolation may lead to incomplete or misleading results, as its execution may depend on a broader calling context, making it necessary to systematically expand the code slice to capture relevant execution paths.
At the same time, expanding too much can introduce unnecessary complexity and computational overhead. 
We balance these tradeoffs by iteratively growing the code slice around the vulnerable function.
This involves two key sub-steps:

\vspace{1mm}
\noindent
\textit{(3a) Define and grow the code slice.}  
To systematically expand the execution context, we need a structured way to group slices based on their size and context. 
We introduce the concept of slice level, which groups slices based on their distance from the vulnerable function in the call graph.
Our slicing process follows these principles:
\begin{itemize}
    \item {Level 0:} Contains only the vulnerable function.
    \item {Level 1:} Adds all immediate callers of the vulnerable function.
    \item {Level 2:} Adds callers of Level 1 functions, etc.
\end{itemize}
   
Each level adds more calling context, allowing us to test whether the vulnerability warning depends on external inputs or conditions. 
If a vulnerability is triggered at Level N, further expansion is necessary as higher levels may reveal hidden preconditions that prevent the vulnerability from being exploited. If the suspected code location is reached sufficiently but no vulnerability is triggered, further expansion may be unnecessary~\cite{FuzzSlice}.

\begin{figure}[tb!]
    \centering {\includegraphics[width=0.48\columnwidth]{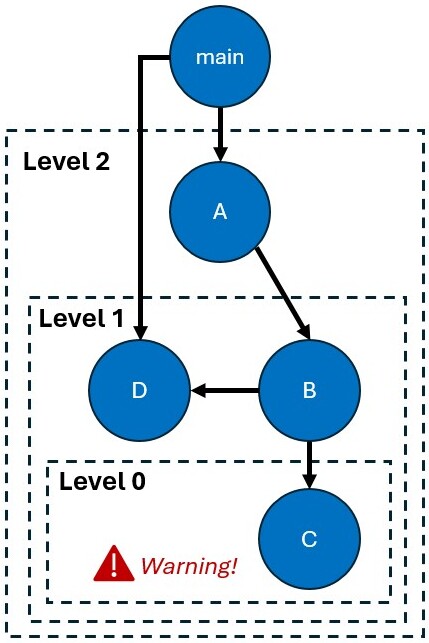}}   
    \caption{An example illustrating slice levels where each node represents a function.}
    \label{img:levels}
\end{figure}

Figure~\ref{img:levels} shows a call graph where \texttt{main} calls both \texttt{A} and \texttt{D}; \texttt{A} calls \texttt{B}, which in turn calls \texttt{C} and \texttt{D}. The vulnerable function is \texttt{C}.
The slice expands progressively, i.e., in Level 0, only function \texttt{C} is included (minimal slice).  
Level 1 includes functions \texttt{B}, \texttt{C}, and \texttt{D} in one slice, as \texttt{B} has a maximum calling distance of 1 from \texttt{C} and \texttt{B}'s dependencies are \texttt{C} and \texttt{D}.
Level 2 further includes function \texttt{A} in a single slice, with a maximum calling distance of 2 from \texttt{C}. The slice includes \texttt{A} and its recursive dependencies \texttt{B}, \texttt{C}, \texttt{D}. Depending on the callgraph, there can be multiple slices at a given level.

\vspace{1mm}
\noindent
\textit{(3b) Compile code slices using \oldfs.}
\textcolor{black}{After \fs selects an entry function for a particular slice level, it invokes \oldfs to compile that function-rooted slice. \oldfs resolves cross-file dependencies, removes code unnecessary for compilation, produces a standalone binary, and generates a baseline fuzzing wrapper. In this architecture, \oldfs serves as a slice-construction component. \oldfs does not construct \fs's static call graph, select caller roots, define slice levels, schedule exploration across levels, aggregate outcomes, or produce final warning-level classifications. These operations belong to \fs's multi-level orchestration and evidence-aggregation methodology. \fs also extends \oldfs's wrapper generation by trimming unused wrapper variables and adding support for fuzzing enums and bitfield data types. These changes improve input efficiency and broaden the set of functions that can be tested.}

For each slice level, we identify all functions whose distance from the vulnerable function matches that level. In Figure~\ref{img:levels}, for instance, only function \texttt{C} (the vulnerable function) is sliced at {Level 0}. 
At {Level 1}, we slice function \texttt{B}, which is one level up from \texttt{C}, and its dependencies.
At {Level 2}, function \texttt{A} is also included.

This step finally produces multiple binaries, each containing a code slice with a defined amount of context.


\vspace{1mm}
\noindent
\textbf{{(4) Fuzz binaries and classify the vulnerability.}}
In this step, we fuzz all generated binaries at each slice level to determine whether the suspected code location is potentially a true bug or a false alarm.

At this stage, multiple code slices have been compiled and linked, each corresponding to a different execution path that leads to the suspected function. 
We iteratively perform fuzzing, starting from binaries at Level 0, followed by those at Level 1, and so forth. \fs uses AFL++~\cite{afl++} to fuzz the binaries and collects coverage information using llvm-coverage\cite{BibEntry2023Jan} in order to track execution frequencies of each line, alongside the results of runtime bug oracles (e.g., ASAN\cite{serebryany2012addresssanitizer} and UBSAN~\cite{UBSAN}) to detect vulnerabilities.

The challenge at this stage is to consolidate the fuzzing results across different execution paths and slice levels to make a final classification of the vulnerability warning.
We perform classification at  three stages:

\noindent
\textit{(1) Classification per binary.} 
Each binary is fuzzed independently, and its behavior is classified into one of three states based on coverage and bug detection results:

\begin{enumerate}[label=(\alph*)]
    \item Triggered: The bug oracle detects a vulnerability at the warning location, confirming that the warning has been triggered in the code slice.
    
    \item Reached: The warning location is executed, but no bug is detected by the bug oracle, suggesting that while the path is feasible, it does not necessarily lead to an actual vulnerability.
    
    \item Not Reachable: The warning location is never executed during fuzzing, indicating that no execution path in the slice reaches it.
\end{enumerate}

\noindent
\textit{(2) Classification per slice level.}
Since multiple binaries may exist at a given slice level--each representing different execution paths--the fuzzing results must be aggregated at this level before making further decisions.
We aggregate the results from all binaries at a given slice level as follows.

\begin{enumerate}[label=(\alph*)]
    \item Vulnerable (V): If at least one binary at the current slice level is triggered, then a vulnerable execution path has been identified. This suggests that the warning location is likely a true positive.

    \item Benign (B): If no slice at the current level is triggered, but at least one slice is reached, it indicates that execution paths to the vulnerability warning exist, but no vulnerabilities have been exploited.
    This suggests that the warning is likely a false positive, as an extensive search fails to identify an offending input that triggers the bug.

    \item Not Reachable (NR): If all binaries at the current level classify the warning as Not Reachable, no feasible execution path leads to the warning at this level. In such cases, the warning location remains undetermined, and we cannot make a conclusive classification at this slice level.
\end{enumerate}

\noindent
\textit{(3) Final classification across all levels.} 
After analyzing individual binaries and consolidating the results at each slice level, the final classification of the warning is determined based on its behavior across multiple levels. 
If the analysis is limited to a predefined number of slice levels (e.g., Level 0 to Level 2), we make the final decision as follows:




\begin{enumerate}[label=(\alph*)]
    \item Possible True Positive (PTP): If the warning is consistently labeled as ``Vulnerable'' at all slice levels, it is classified as a PTP, indicating a high likelihood of an actual vulnerability.

    \item Possible False Positive (PFP): If the warning is reached through ``Benign'' paths at lower slice levels and remains benign or ``Not Reachable'' at higher levels, it is classified as a PFP.
    This reflects our validated assumption (Section~\ref{sec:results}) that warnings benign at shallow slices are unlikely to become vulnerable in deeper ones~\cite{FuzzSlice}.

    \item Not Reachable (NR): If the warning remains ``Not Reachable'' at all slice levels, it means that conclusions cannot be drawn about the validity of the warning, as it was never reached during fuzzing.
\end{enumerate}

The final classification rule is a heuristic evidence-aggregation strategy rather than a formal proof of vulnerability or benignness. Its purpose is to support warning triage by summarizing how a warning behaves across increasingly broader sliced execution contexts. Consistent triggering across slice levels increases confidence that the warning should be prioritized, while reaching the warning without triggering the bug oracle provides evidence that the warning may be benign under the explored contexts. Accordingly, we use cautious labels—Possible True Positive, Possible False Positive, and Not Reachable—to reflect the evidential nature of the classification.

\section{\fs Optimizations}
\label{sec:optimizations}
While \fs systematically tests vulnerability warnings through iterative slicing and fuzzing, practical challenges may arise. 
Specifically, as slice levels grow, the number of execution paths can increase exponentially, leading to excessive slice creation and redundant fuzzing. 
Since each execution path requires a separate binary for analysis, inefficient slicing and fuzzing can significantly inflate computational costs.

To address these challenges, \fs proposes a set of optimizations that fall into two categories: (1) Slice Creation Optimizations; (2) Slice Fuzzing Optimizations.
Below, we describe each set of optimizations in detail.

\subsection{Slice Creation Optimizations}
These optimizations focus on reducing time to construct slices, especially when slices overlap across levels.


\vspace{1mm}
\noindent
\textbf{(O1) Cache minimized slices during recursive slicing.}
When slicing a function, \fs recursively minimizes its dependencies to construct a compilable unit.  
At higher slice levels, these dependencies often overlap across different slices.  
To eliminate redundant work, \fs caches minimized slices for individual dependencies.  
When a previously minimized dependency is needed again, it is retrieved directly from the cache, avoiding re-minimization and recompilation.
This also helps prevent redundant compilation when there are cyclic function calls in the project.
This reduces the cost of constructing large slices at higher levels.

\vspace{1mm}
\noindent
\textbf{(O2) Parallelize slice creation.} Slice creation can be efficiently parallelized across multiple cores since it involves copying files from the project and trimming code extraneous to each slice, followed by compiling the slice~\cite{FuzzSlice}. Workloads are distributed across multiple cores with centralized slice caching.

\subsection{Slice Fuzzing Optimizations}
These optimizations focus on improving the efficiency of fuzzing once the slices have been constructed.

\vspace{1mm}
\noindent
\textbf{(O3) Skip redundant fuzzing after triggering a vulnerable path.}  
During fuzzing at a given slice level, \fs generates multiple binaries corresponding to different ancestor functions.  
Once fuzzing triggers the vulnerability in any binary, further fuzzing at that level becomes redundant for classification purposes.
Thus, \fs halts fuzzing remaining slices at that level immediately after the first successful vulnerability exploit.

\vspace{1mm}
\noindent
\textbf{(O4) Skip fuzzing at higher levels for non-triggering warnings.}
If slices at a given level reach the warning location but do not trigger the bug oracle, \fs skips fuzzing higher-level slices for that warning.
This optimization acts as an early stopping rule for this specific case, i.e., once a warning has been exercised without triggering a failure, additional caller context is unlikely to make the same warning vulnerable under our validation model. 
This optimization does not define the full stopping condition of \fs, but it implements one stopping decision used to avoid redundant higher-level fuzzing.

\vspace{1mm}
\noindent
\textbf{(O5) Parallelize fuzzing of independent slices.}  
Each slice results in a single standalone binary, independent of other slices. 
\fs exploits this by enabling parallel fuzzing of slices across cores or machines without requiring synchronization.
Unlike full-program fuzzing, which needs complex input sharing across threads, fuzzing independent slices achieves near-linear scalability, hence reducing total analysis time.

\vspace{1mm}
\noindent
\textbf{(O6) Remove code irrelevant to warning execution.} 
To accelerate fuzzing, \fs prunes code that cannot influence the vulnerability’s execution.  
Using static intra-procedural AST analysis, \fs removes code appearing after the vulnerability location that is not enclosed within control-flow structures (e.g., \texttt{if} or \texttt{while} blocks).  
This minimization can reduce fuzzing overhead and improve the probability of rapidly triggering the vulnerability.
This pruning step is intra-procedural and hence unaffected by cycles in the call graph.
In the case of higher slice levels, we prune the entry-point function by removing code appearing after the last vulnerable function call leading to the vulnerability.

\section{Evaluation}
\label{sec:results}
We organize the evaluation of \fs around the following research questions:


\begin{itemize}
\item \noindent
\textbf{RQ1:} {\RQone}

\item \textbf{RQ2:} {\RQtwo{}}

\item \textbf{RQ3:} {\RQthree{}}

\item \textbf{RQ4:} {\RQfour{}} 

\item \textbf{RQ5:} {\RQfive{}}

\item \textbf{RQ6:} {\RQsix{}} 

\item \textbf{RQ7:} {\RQseven{}} 




\end{itemize}


To answer these questions, we first describe our evaluation setup in Section~\ref{setup}, including dataset construction and \fs configuration. We then report results for each RQ.
in Sections~\ref{res1}–\ref{res7}.



\subsection{Evaluation Setup}
\label{setup}
In this section, we describe the datasets, the environment, and the configuration used to evaluate \fs.

\vspace{1mm}
\noindent 
\textbf{Dataset.}
\fs is designed to validate individual code locations, such as those emerged by static analyzers or introduced in recent commits, by slicing and fuzzing the surrounding program context. 
To evaluate \fs, we require datasets with ground truth: confirmed true vulnerabilities and false alarms.
%

We use the RevBugbench framework~\cite{Zhang2022} to inject true vulnerabilities into three real-world projects: \href{https://github.com/GNOME/libxml2}{libxml2}, \href{https://github.com/facebook/zstd}{zstd}, and \href{https://github.com/OSGeo/PROJ}{PROJ}.
RevBugbench accepts a bug fix pattern as input, specifying both the syntactic structure and semantic conditions of a bug. 
If a code location matches this pattern, RevBugbench ``reverts'' the fix, reintroducing a likely bug into the code. The resulting buggy version contains a true bug in the ``reverted'' function, while a corresponding warning of the same function in the fixed version (i.e., without the reverted fix) serves as a false alarm.

We validate the behavior of each injected and fixed variant by running existing test suites of the project, confirming that: (1) the injected version causes a failure, and (2) the fixed version passes the test suite without triggering the bug. 
This allows us to create two sets of suspicious code locations per project: one with confirmed vulnerabilities and one with confirmed false alarms.

Table~\ref{Repo stats} provides statistics on the projects in our benchmark.
Of the three projects, zstd contains the largest number of vulnerabilities (49), followed by libxml2 (38), and PROJ (10).
They include 20 heap overflow, 59 stack overflow, 3 floating-point exceptions, 5 global overflows, 8 memory leaks, and 2 stack use-after-scope vulnerabilities.
The 97 true vulnerabilities forms \textbf{dataset-TP} and 97 false alarms forms \textbf{dataset-FP}.
Both datasets form the foundation for all subsequent experiments.

We use RevBugbench for realistic evaluations, as it preserves full call hierarchies in real-world projects by injecting bugs, in contrast to synthetic benchmarks like Juliet~\cite{juliet}. On the other hand, CVE-based datasets allow evaluating multiple real vulnerabilities but require switching artifacts across project versions. RevBugbench avoids this, enabling consistent slicing and fair evaluation with directed fuzzing all on the same project version.


\vspace{1mm}
\noindent
\textbf{Configuration.}  
We evaluate \fs on each examined code location using iterative slicing and fuzzing at \textit{three} slice levels, with time budgets increasing by complexity: 5, 7.5, and 10 minutes for Level 0, 1, and 2, respectively.
We limited experiments to Level 2 due to practical scalability constraints, not because detection plateaued. Higher slice levels cause a rapid growth in the number of slices, substantially increasing fuzzing time.
Experiments are first run without optimizations (for RQ1--RQ3) to assess classification coverage, then repeated with optimizations (for RQ4--RQ5) to measure time savings, without affecting accuracy.
A bare-metal Kubernetes cluster with two AMD EPYC 9224 (48-core, 2.5 GHz) worker nodes, each with 1 TB RAM, is used to execute the experiments. Fuzzing jobs are pinned to a single core and allocated 20 GB RAM, with a maximum of 20 jobs running concurrently.

\begin{table}[tb]
  \scriptsize
  \caption{Statistics on the repositories in our benchmark.}
  \centering
  \setlength{\tabcolsep}{3pt}
  \renewcommand{\arraystretch}{1.1}
  \begin{tabular}{lrr}
    \toprule
    \textbf{Repository} & \textbf{Dataset-TP (\#True Vulns.)} & \textbf{Dataset-FP (\#False Alarms)} \\
    \midrule
    libxml2 & 38 & 38 \\
    zstd    & 49 & 49 \\
    PROJ    & 10 & 10 \\
    \bottomrule
  \end{tabular}
  \label{Repo stats}
\end{table}

\subsection{RQ1: \RQone}
\label{res1}

\noindent
\textit{\underline{Goal and Approach}}. RQ1 evaluates whether \fs can trigger true vulnerabilities across varying slice levels.
We evaluate this by applying \fs to dataset-TP.
For each vulnerability location, we generate slices at three levels (Level 0, 1, 2), incrementally expanding the surrounding code context. 
At each slice level, we classify the outcome as (1) Vulnerable (V): bug is triggered, (2) Benign (B): location reached but no vulnerability triggered, or (3) Not Reachable (NR): location not reached.

\begin{table}[tb!]
  \scriptsize
  \caption{
    \fs classification outcomes for true vulnerabilities across three slice levels. 
    \textbf{V} = Bug oracle is triggered at the location in at least one slice at the level, 
    \textbf{B} =  Location is reached in at least one slice but the bug oracle is not alerted,
    \textbf{NR} = Not Reachable (location unreached across slice level), 
    \textbf{PTP} = possible true positive, 
    \textbf{PFP} = possible false positive.
    The locations consistently crashing across all slice levels are only considered as possible true positives.
  }
  \centering
  \setlength{\tabcolsep}{3pt}
  \renewcommand{\arraystretch}{1.1}
  \resizebox{\columnwidth}{!}{
  \begin{tabular}{lcrrrrccc}
    \toprule
    \textbf{Project} & \textbf{\# Vulns.} & \textbf{Lvl} & \textbf{V} & 
    \textbf{B} & \textbf{NR} & \textbf{PTP} & \textbf{PFP} & \textbf{NR}\\
    \midrule
    \multirow{3}{*}{libxml2} 
      & \multirow{3}{*}{38} & 0 & 30 & 0 & 8  & \multirow{3}{*}{20} & \multirow{3}{*}{0} & \multirow{3}{*}{18} \\
      &                     & 1 & 26 & 0 & 12 &                      &                     \\
      &                     & 2 & 20 & 0 & 18 &                      &                     \\
    \midrule
    \multirow{3}{*}{zstd} 
      & \multirow{3}{*}{49} & 0 & 47 & 0 & 2  & \multirow{3}{*}{29} & \multirow{3}{*}{0} & \multirow{3}{*}{20} \\
      &                     & 1 & 38 & 0 & 11 &                      &                     \\
      &                     & 2 & 29 & 0 & 20 &                      &                     \\
    \midrule
    \multirow{3}{*}{PROJ} 
      & \multirow{3}{*}{10} & 0 &  9 & 0 & 1  & \multirow{3}{*}{4}  & \multirow{3}{*}{0} & \multirow{3}{*}{6} \\
      &                     & 1 &  7 & 0 & 3  &                      &                     \\
      &                     & 2 &  4 & 0 & 6  &                      &                     \\
    \midrule
    \textbf{Total} & \textbf{97} & -- & -- & -- & -- & \textbf{53 (54.6\%)} & \textbf{0 (0\%)} & \textbf{44 (45.3\%)}\\
    \bottomrule
  \end{tabular}
  }
  \label{truebugs}
\end{table}
\vspace{0.5ex}
\noindent
\textit{\underline{Results}}.
Table~\ref{truebugs} summarizes the classification results of the examined vulnerable code locations. 
\textcolor{black}{From the table, we observe that \textbf{\fs triggers the bug oracle consistently across all analyzed levels for 53 of 97 confirmed vulnerabilities (54.6\%), without any misclassifications.} }
That is, \fs correctly triggers the vulnerability at all three slice levels for 53 vulnerable code locations (54.6\%), including 20/38 in {libxml2}, 29/49 in {zstd}, and 4/10 in {PROJ}.  
In all cases where a vulnerability location is reachable, the bug is also triggered, resulting in no misclassifications.
 
We also observe that vulnerability reachability declines at higher slice levels due to increased complexity. For example, at Level~2, 33 vulnerable code locations (34\%) that were reachable and triggered at Level~0 become Not Reachable.  
This drop is due to added dependencies in deeper slices, such as function pointers, environment-specific conditions, and file system interactions, which complicate fuzzing and require longer exploration time.
For projects such as PROJ, a larger fraction of injected locations becomes dependent on function pointers as the slice expands, causing more Level 2 slices to be classified as NR.

\textbf{Expanding slicing scope might trigger previously unreachable vulnerabilities.}
Although larger slices can complicate fuzzing, they can also expose vulnerabilities that were unreachable in smaller contexts. 
In 3.1\% of the examined cases, we find that vulnerable code locations that were Not Reachable at lower slice levels became reachable and triggered at higher levels.
These cases highlight the importance of analyzing multiple slice levels to uncover vulnerabilities that depend on a broader initialization or environmental setup.

\begin{figure}[tb!] 
    \centering
    \includegraphics[width=0.5\textwidth]{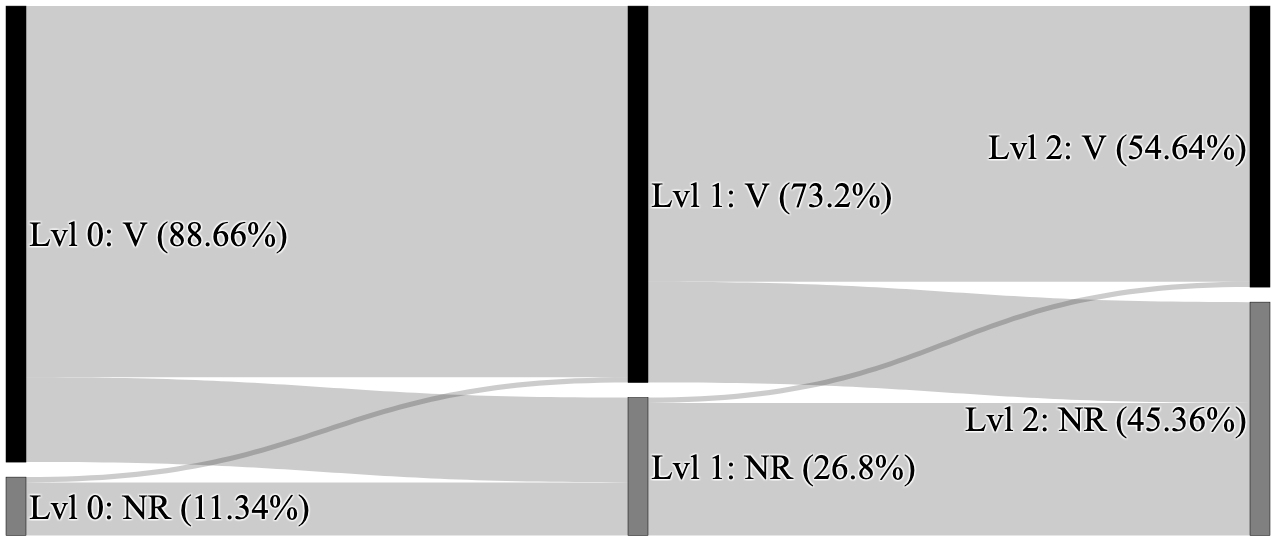}
   \caption{Sankey diagram showing classification transitions of true vulnerabilities across slice levels on dataset-TP. 
   }
    \label{injected-sankey}
\end{figure}

\textbf{Across all slice levels, no vulnerability is observed to shift from vulnerable to benign behavior.}
The Sankey diagram in Figure~\ref{injected-sankey} visualizes classification transitions across levels.  
Among all true vulnerabilities, we find that no bug that was initially classified as vulnerable ever transitions to being classified as benign at higher levels.  
\textcolor{black}{This observed stability supports the empirical basis of \fs cross-level aggregation in our benchmark.}

\subsection{RQ2:~\RQtwo}
\label{res2}
\noindent
\textit{\underline{Goal and Approach}}.
While RQ1 focused on the ability of \fs to detect true vulnerabilities, RQ2 complements the analysis by measuring its effectiveness in eliminating false alarms, which is crucial for prioritizing bug reports to triage.
As before, we analyze how these false alarms behave across increasing slice levels.

\begin{table}[tb!]
  \scriptsize
  \caption{
    \fs classification outcomes for false alarms across three slice levels. 
    \textbf{V} = Bug oracle is triggered at the  location in at least one slice at the level, 
    \textbf{B} = Location is reached in at least one slice but the bug oracle is not alerted,
    \textbf{NR} = Not Reachable (location unreached across slice level), 
    \textbf{PTP} = possible true positive, 
    \textbf{PFP} = possible false positive.
    The locations that are benign at lower levels and unreachable or benign at higher levels are tallied together to obtain possible false positives.
  }
  \centering
  \setlength{\tabcolsep}{2pt}
  \renewcommand{\arraystretch}{1.1}
  \resizebox{\columnwidth}{!}{
  \begin{tabular}{lcrrrrccc}
    \toprule
    \textbf{Repository} & \textbf{\# Vulns.} & \textbf{Lvl} & \textbf{V} & 
    \textbf{B} & \textbf{NR} & \textbf{PTP} & \textbf{PFP} & \textbf{NR} \\
    \midrule
    \multirow{3}{*}{libxml2}
      & \multirow{3}{*}{38} & 0 & 14 & 17 &  7 & \multirow{3}{*}{7}  & \multirow{3}{*}{21} & \multirow{3}{*}{10} \\
      &                     & 1 & 11 & 14 & 13 &                     &                      \\
      &                     & 2 &  7 & 12 & 19 &                     &                      \\
    \midrule
    \multirow{3}{*}{zstd}
      & \multirow{3}{*}{49} & 0 & 30 & 13 &  6 & \multirow{3}{*}{20} & \multirow{3}{*}{26} & \multirow{3}{*}{3} \\
      &                     & 1 & 24 & 18 &  7 &                     &                      \\
      &                     & 2 & 20 & 26 &  3 &                     &                      \\
    \midrule
    \multirow{3}{*}{PROJ}
      & \multirow{3}{*}{10} & 0 &  3 &  6 &  1 & \multirow{3}{*}{1}  & \multirow{3}{*}{7}  & \multirow{3}{*}{2} \\
      &                     & 1 &  1 &  5 &  4 &                     &                      \\
      &                     & 2 &  1 &  0 &  9 &                     &                      \\
    \midrule
    \textbf{Total} & \textbf{97} & -- & -- & -- & -- & \textbf{28 (28.8\%)} & \textbf{54 (55.6\%)} & \textbf{15 (15.4\%)} \\
    \bottomrule
  \end{tabular}
  }
  \label{fixedbugs}
\end{table}

\begin{figure}[tb!] 
    \centering 
    \includegraphics[width=0.5\textwidth]{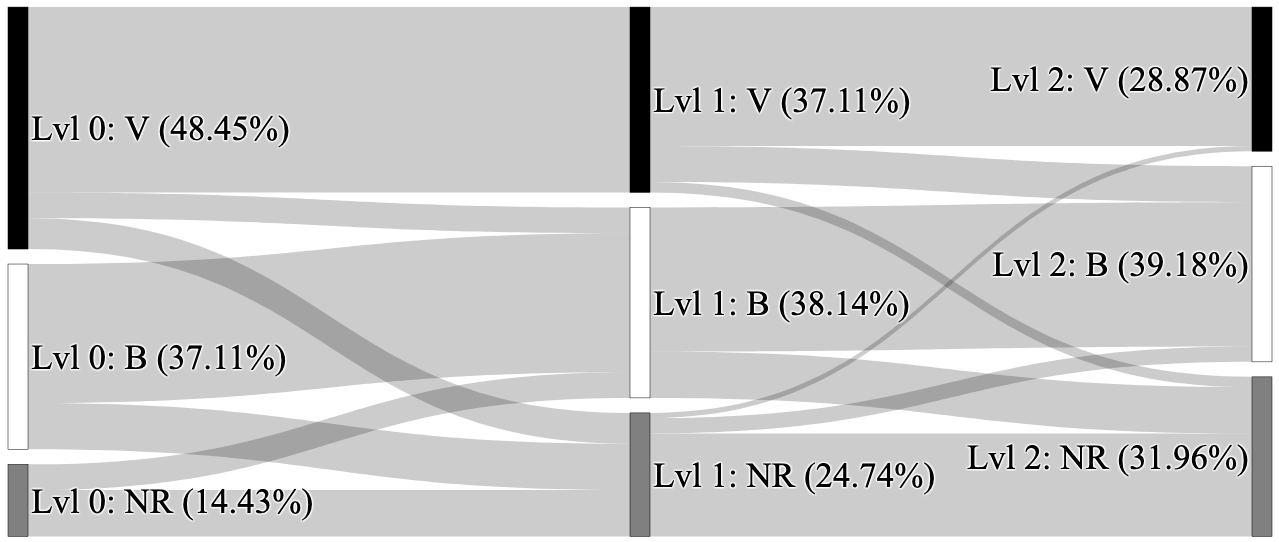} 
    \caption{Sankey diagram showing classification transitions of false-alarm locations across slice levels on dataset-FP.} 
    \label{notinjected-sankey} 
\end{figure}
\vspace{0.5ex}
\noindent
\textit{\underline{Results}}.
As we can see from Table~\ref{fixedbugs}, across the 97 false alarms in our dataset-FP, \fs classifies 54 locations (55.6\%) as Possible False Positives (PFP), indicating that only benign execution paths are observed across slice levels 0--2.
\textbf{\textcolor{black}{\fs produces Possible False Positive evidence for a considerable portion of confirmed false alarms at minimal slice levels.}
}
We note that 36 out of these 54 cases (37.1\% of all locations) are already classified as benign at level 0, without requiring expansion to larger slices.
This shows that \fs can provide early prioritization evidence without expanding every warning to higher levels.

\textbf{Benign classifications at lower levels remain stable or conservative at higher levels.}
Figure~\ref{notinjected-sankey} visualizes how location classifications evolve across slice levels.
\textcolor{black}{In our experiments, locations classified as benign at lower levels (B) did not subsequently exhibit vulnerable behavior (V) at higher levels.
The observed transitions support using lower-level benign outcomes as heuristic evidence for being a Possible False Positive.}
We do observe a small number of locations (16) that transition from benign (B) to unreachable (NR) at level 2, mainly due to added environmental complexity in larger slices (e.g., file I/O dependencies).
However, conservatively, we classify these cases as PFP because they had previously demonstrated benign behavior when reached in smaller slices.
For example in the case of PROJ, we have 7 PFPs which arise from 6 benign locations at Level 0 and an additional benign location at Level 1 (previously vulnerable at Level 0). All of these become unreachable at Level 2 but still are considered as a Possible False Positive due to benign behaviour in smaller slices.

\textbf{Incomplete slice context can occasionally mislead classification, but higher levels improve classification accuracy.}
From Table~\ref{fixedbugs}, we also observe that 28 suspicious code locations (28.8\%) are misclassified as vulnerable (PTP) across all levels, despite belonging to the false alarms dataset-FP. 
This misclassification is primarily attributed to the Fixreverter framework~\cite{Zhang2022}, where injected bugs alter guard conditions outside the slice context.
While expanding slices beyond level 2 could potentially resolve these cases, it would introduce significant computational cost.
Conversely, we observe that 12 false vulnerabilities initially showing vulnerable behavior at lower levels are correctly reclassified as benign at higher levels (5 at level 1 and 7 at level 2), indicating that moderate slice expansion can recover correct classification.


\subsection{RQ3: \RQthree}
\label{res3}

\noindent
\textit{\underline{Goal and Approach}}.
Beyond detecting vulnerabilities or filtering false alarms, the focus of RQ3 is on fidelity--whether \fs triggers vulnerabilities through realistic calling contexts.
This fidelity is crucial for assessing the semantic relevance of the vulnerability trigger.  
Since many vulnerabilities reside deep in nested call chains, reproducing a vulnerability from a function in the stack trace of an exploit suggests that \fs follows a meaningful path.
Such fidelity not only validates the detection but also helps developers understand the bug’s triggering context, potential exploit vectors, and severity. 
Moreover, \fs may uncover alternate calling paths not present in the original trace, revealing new avenues for exploitation.

To evaluate this, for each true vulnerability in dataset-TP, we compare the entry function of the slice fuzzed at level~$k$ with the $k^\text{th}$ function in the stack trace generated by executing the test suite on the buggy version of the project. 
We count a match when \fs triggers the vulnerability from a slice rooted at a function that appears in the actual bug stack trace. 

Listing~\ref{BugStacktrace} illustrates the crash stack trace of a heap buffer overflow from the {libxml2} project:


\begin{lstlisting}[language=C++, 
  caption={Bug stacktrace of a heap buffer overflow.}, 
  label={BugStacktrace},
  escapeinside={(*@}{@*)},
  basicstyle=\ttfamily\scriptsize,
  columns=flexible,
  xleftmargin=0.7em,
  framexleftmargin=0.7em,
  framexrightmargin=0em,
]
(*@\textcolor{red}{\#0}@*) xmlGetPredefinedEntity  /libxml2/entities.c:496:13
(*@\textcolor{red}{\#1}@*) xmlSAX2GetEntity        /libxml2/SAX2.c:872:8
(*@\textcolor{red}{\#2}@*) xmlParseEntityRef       /libxml2/parser.c:9740:12
(*@\textcolor{red}{\#3}@*) LLVMFuzzerTestOneInput  /libxml2/fuzzer.cc:87:10
\end{lstlisting}

Although the vulnerability lies in \texttt{xmlGetPredefinedEntity}, it is only reachable via multiple intermediate calls. 
To test whether \fs replicates this path, we fuzz slices rooted at higher-level functions. 
For example, a Level~1 slice rooted at \texttt{xmlSAX2GetEntity} may trigger the bug and thus replicate the second frame in the trace. 
Similarly, a Level~2 slice rooted at \texttt{xmlParseEntityRef} may reproduce the bug from the third stack frame.
These cases indicate that \fs can successfully trace backward and isolate vulnerable paths that align with real crash behavior by only fuzzing partial code slices.

To obtain the ground truth, we use the project’s existing test suite to execute the injected vulnerability, then collect the resulting stack trace using dynamic sanitizers such as ASAN~\cite{serebryany2012addresssanitizer} and UBSAN~\cite{UBSAN}.
We then compare this against the stack trace reported by \fs when it triggers a vulnerability. 
We specifically compare the top three frames to determine if the slice entry function corresponds to one of the real call frames. We also ensure that the type of vulnerability reported (e.g., heap overflow, use-after-free) is consistent between both traces.

\begin{table}[tb]
  \scriptsize
  \caption{
    Alignment between ground-truth exploit stack traces and vulnerable paths identified by \fs. 
    \textbf{V} = Bug oracle alerted at the vulnerable location in at least one slice.
  }
  \centering
  \setlength{\tabcolsep}{10pt}
  \renewcommand{\arraystretch}{1.1}
  \begin{tabular}{llrrr}
    \toprule
    \textbf{Repository} & \textbf{Level} & \textbf{\# True Vulns.} & \textbf{V} & \textbf{Aligns} \\
    \midrule
    \multirow{3}{*}{libxml2}
      & 0 & 38 & 30 & 30 (78.9\%) \\
      & 1 & 38 & 26 & 19 (50.0\%) \\
      & 2 & 38 & 20 & 11 (28.9\%) \\
    \midrule
    \multirow{3}{*}{zstd}
      & 0 & 49 & 47 & 47 (95.9\%) \\
      & 1 & 49 & 38 & 31 (63.2\%) \\
      & 2 & 49 & 29 & 25 (51.0\%) \\
    \midrule
    \multirow{3}{*}{PROJ}
      & 0 & 10 &  9 &  9 (90.0\%) \\
      & 1 & 10 &  7 &  5 (50.0\%) \\
      & 2 & 10 &  4 &  3 (30.0\%) \\
    \bottomrule
  \end{tabular}
  \label{stacktracecomp}
\end{table}

\vspace{1ex}
\noindent
\textit{\underline{Results}}. 
Table~\ref{stacktracecomp} presents the evaluation results for RQ3.  
For each project, the table reports: (1) the number of true vulnerabilities with available ground-truth stack traces, (2) the number of cases where location exhibits vulnerable behavior (V) at given slice level, and (3) the number of cases where the slice entry function aligns with a function in the ground-truth stack trace (Aligns). 
Below, we present the main observations.

\textbf{Tight alignment at shallow slices.} 
From Table~\ref{stacktracecomp}, we observe that at Level~0, \fs replicates the ground-truth stack trace in 88.6\% of cases across all projects (on average).
Alignment is particularly strong for {zstd} (95.9\%) and {PROJ} (90\%).  
This confirms that small, minimal slices often preserve enough execution context to detect true vulnerabilities.

\textbf{Alignment decreases as slice levels increase.}
We also observe that the average alignment drops from 88.6\% at Level~0 to 56.7\% at Level~1 and 40.2\% at Level~2, due to increased slice complexity, e.g., added control flow, environmental dependencies, and stricter input constraints, making it harder to reproduce ground-truth stack traces during fuzzing.
Additionally, since \fs relies on a static call graph, it may miss dynamic edges (e.g., indirect calls), leading to deviations from actual call paths.  
Still, \fs often triggers vulnerabilities through alternative paths, maintaining detection effectiveness despite imperfect alignment.


\subsection{RQ4: \RQfour}
 \label{res4}

\begin{table}[tb]
  \scriptsize
  \caption{
    Effect of optimizations (O1-O5) on analysis-time across Levels 0-2 in \fs. Times are reported in days.
  }
  \centering
  \setlength{\tabcolsep}{6.9pt}
  \renewcommand{\arraystretch}{1.1}
  \begin{tabular}{l c rr rr}
    \toprule
    \multirow{2}{*}{Project} & \multirow{2}{*}{Dataset} &
    \multicolumn{2}{c}{\textbf{No Optimizations}} &
    \multicolumn{2}{c}{\textbf{With Optimizations}} \\
    \cmidrule(lr){3-4} \cmidrule(lr){5-6}
    & & Slice & Fuzz & Slice & Fuzz \\
    \midrule
    \multirow{2}{*}{libxml2}
      & True Vulns.  & 0.650 & 23.370 & 0.032 & 0.644 \\
      & False Alarms & 0.650 & 23.370 & 0.032 & 0.535 \\
    \midrule
    \multirow{2}{*}{zstd}
      & True Vulns.  & 0.021 & 5.598  & 0.003 & 0.284 \\
      & False Alarms & 0.021 & 5.598  & 0.003 & 0.223 \\
    \midrule
    \multirow{2}{*}{PROJ}
      & True Vulns.  & 0.096 & 0.489 & 0.019 & 0.179 \\
      & False Alarms & 0.096 & 0.489 & 0.019 & 0.152 \\
    \midrule
    \textbf{Total} & -- & \textbf{1.534} & \textbf{58.914} & \textbf{0.108} & \textbf{2.017} \\
    \bottomrule
  \end{tabular}
  \label{time-optimizations}
\end{table}


\noindent
\textit{\underline{Goal and Approach}}.
This RQ examines how the optimizations described in Section~\ref{sec:optimizations} impact the overall analysis time of \fs, aiming to improve its scalability without sacrificing classification accuracy.  
Unlike previous RQs, where \fs was run without optimizations to measure baseline effectiveness, this RQ focuses on practical efficiency.

We re-run \fs on the vulnerability dataset-TP and dataset-FP while enabling all optimizations (O1–O6), including slice caching,  fuzzing termination heuristics, and parallelization across 20 CPU cores.  
We separately measure (1) slice creation time, (2) fuzzing time, and (3) the number of slices analyzed, and compare them against the unoptimized single core baseline of \fs.

\begin{table}[tb!]
  \scriptsize
  \caption{
  Effect of optimizations (O3-O4) on number of slices fuzzed by \fs. 
  \textbf{\%} = \% of slices skipped for fuzzing due to optimizations.
  \emph{Level 0 results are omitted as they are unaffected by fuzzing optimizations (O3-O4).}
}
  \centering
  \setlength{\tabcolsep}{5pt}
  \renewcommand{\arraystretch}{1.15}
  \begin{tabular}{l c rrr rrr}
    \toprule
    \multirow{2}{*}{Project} & \multirow{2}{*}{Dataset} & 
    \multicolumn{3}{c}{Level 1} & \multicolumn{3}{c}{Level 2} \\
    \cmidrule(lr){3-5} \cmidrule(lr){6-8}
    & & Before & After & \% & Before & After & \% \\
    \midrule
    libxml2 & True Vulns.  & 694 & 208 & 70.0 & 1736 & 837 & 51.7 \\
            & False Alarms & 694 &  71 & 89.7 & 1736 & 182 & 89.5 \\
    \midrule
    zstd    & True Vulns.  & 102 & 66 & 35.2 &  162 &  86 & 46.9 \\
            & False Alarms & 102 & 42 & 58.8 &  162 &  50 & 69.1 \\
    \midrule
    PROJ    & True Vulns.  &  25 &  18 & 28.0 &   26 &  24 &  7.6 \\
            & False Alarms &  25 &  13 & 48.0 &   26 &   9 & 65.3 \\
    \bottomrule
  \end{tabular}
  \label{ancestor-optimizations}
\end{table}

\vspace{1ex}
\noindent
\textit{\underline{Results}}. Table~\ref{time-optimizations} discusses the effect of optimizations on \fs analysis time. \textbf{\fs slice optimizations (O1-O2) reduce slice creation time by 92.9\% on average across projects.}  
\textbf{Additionally, \fs fuzzing optimizations accelerated by parallelization (O3-O5) reduce fuzzing time by 96.5\% across projects.}  

Pruning code unnecessary for vulnerability execution (O6) improves reachability without directly affecting runtime. Though it doesn't reduce slicing or fuzzing time, it increases the likelihood of reaching vulnerabilities during fuzzing, yielding a 35\% increase in vulnerabilities reached across slice levels in our experiments.

\textbf{Optimizations reduce the number of slices fuzzed by 70.7\% on average across projects.}  
Table~\ref{ancestor-optimizations} shows that optimizations (O3-O4) reduce the number of slices fuzzed by 70.7\% overall.  
Specifically, slice analysis is optimized by 54.8\% on dataset-TP and achieves an 86.6\% reduction on dataset-FP.
This trend confirms that skipping higher-level slice analysis (O4) is more effective on false alarms which are not triggered at shallow levels.


\subsection{RQ5: \RQfive}
\label{res5}

\noindent
\textit{\underline{Goal and Approach}}.
This RQ compares \fs against distance-based directed fuzzer AFLGo in terms of (1) code reachability and (2) analysis time.  
AFLGo~\cite{bohme2017directed}, operates on the full program using coverage-guided mutations to reach targets, while \fs tests isolated slices around the targets only.
We compare \fs with AFLGo, a commonly adopted baseline in comparative studies~\cite{Chen2018Oct, Huang2022May, 272157}, to improve comparability with prior results.

We compare \fs with AFLGo configured in two modes: unseeded (UDF) and seeded (SDF) fuzzing.
The seeded mode uses curated test cases from the project's test suite, excluding inputs that immediately crash or hit targets. The test cases in these projects are extensive, reflecting substantial manual effort to test program features, which provides an advantage to the fuzzer. \fs instead performs fuzzing without seeds across all slices.
All fuzzers are given a 7-day budget and 20 cores, matching the parallel setup of \fs.

For \fs, we enable all optimizations and record the total time to analyze all slices. 
For AFLGo, we report the time to trigger the final unique vulnerability; if no vulnerabilities are triggered, the full 7 day budget is recorded. 
This setup ensures a fair comparison of how quickly each approach produces results.

\vspace{1mm}
\noindent
\textit{\underline{Results}}. \textbf{\fs validates more code locations than unseeded directed fuzzers.}
As shown in Table~\ref{table:timing_totals}, \fs validates 107 target locations, outperforming the unseeded directed fuzzer (40) and closely matching the seeded fuzzer (112).
On large and input-constrained projects such as \texttt{zstd}, the unseeded fuzzer fails to reach any true or false vulnerability locations, whereas \fs validates 29 true vulnerabilities and 26 false alarms respectively, which demonstrates that \fs's slice-based strategy is less affected by complex program entry points.

\textbf{\fs is substantially faster in efficiency than both seeded and unseeded directed fuzzers.}  
\fs completes its entire analysis in 2.09 days across all datasets, compared to 22.23 days for UDF and 11.68 days for SDF, i.e,  \fs achieves a speedup of $10.6\times$ over the unseeded directed fuzzer and $5.5\times$ over the seeded directed fuzzer.
This highlights the efficiency of \fs's parallelized and targeted slice-based analysis.

\textbf{\fs uniquely validates vulnerabilities that directed fuzzers miss.}
Table~\ref{table:intersection_comparison} shows that \fs uniquely validates 86 cases missed by the unseeded fuzzer and 51 missed by the seeded fuzzer, thanks to its ability to analyze localized code without complex runtime setup.
While seeded fuzzer uniquely validates 56 cases, these often depend on global state (e.g., file I/O or environment variables), which is harder to simulate in isolated slices.


\begin{table}[tb]
  \scriptsize
  \caption{
    Comparison of \fs (ST), unseeded directed fuzzing (UDF), and seeded directed fuzzing (SDF). 
    \textbf{Val.} = Validated cases; \textbf{TT} = total time (days).
  }
  \centering
  \setlength{\tabcolsep}{2.5pt}
  \renewcommand{\arraystretch}{1.1}
  \begin{tabularx}{0.49\textwidth}{llr *{6}{r}}
    \toprule
    \multirow{2}{*}{Project} & \multirow{2}{*}{Dataset} & \multirow{2}{*}{\# Cases} &
    \multicolumn{2}{c}{\textbf{\fs (ST)}} &
    \multicolumn{2}{c}{\textbf{UDF}} &
    \multicolumn{2}{c}{\textbf{SDF}} \\
    \cmidrule(lr){4-5} \cmidrule(lr){6-7} \cmidrule(lr){8-9}
    & & & Val. & TT & Val. & TT & Val. & TT \\
    \midrule
    \multirow{2}{*}{libxml2}
      & True Vulns.  & 38 & 20 & 0.67 & 18 & 2.50 & 24 & 2.16 \\
      & False Alarms & 38 & 21 & 0.56 & 20 & 2.63 & 24 & 2.22 \\
    \midrule
    \multirow{2}{*}{zstd}
      & True Vulns.  & 49 & 29 & 0.28 &  0 & 7.00 & 33 & 2.20 \\
      & False Alarms & 49 & 26 & 0.22 &  0 & 7.00 & 28 & 2.63 \\
    \midrule
    \multirow{2}{*}{PROJ}
      & True Vulns.  & 10 &  4 & 0.19 &  1 & 1.43 &  1 & 1.19 \\
      & False Alarms & 10 &  7 & 0.17 &  1 & 1.67 &  2 & 1.28 \\
    \midrule
    \textbf{Total} & -- & \textbf{194} & \textbf{107} & \textbf{2.09} & \textbf{40} & \textbf{22.23} & \textbf{112} & \textbf{11.68} \\
    \bottomrule
  \end{tabularx}
  \label{table:timing_totals}
\end{table}


\begin{table}[tb]
  \scriptsize
  \caption{
    Overlap of validated cases between \fs (ST), the seeded directed fuzzer (SDF) and unseeded directed fuzzer (UDF).
  }
  \centering
  \setlength{\tabcolsep}{1.6pt}
  \renewcommand{\arraystretch}{1.1}
  \begin{tabular}{llr *{6}{r}}
    \toprule
    \multirow{2}{*}{Project} & \multirow{2}{*}{Dataset} & \multirow{2}{*}{\# Cases} &
    \multicolumn{3}{c}{\textbf{\fs vs. UDF}} &
    \multicolumn{3}{c}{\textbf{\fs vs. SDF}} \\
    \cmidrule(lr){4-6} \cmidrule(lr){7-9}
    & & & \makecell[c]{Both\\ST $\cap$ UDF} & \makecell[c]{Only\\ST} & \makecell[c]{Only\\UDF} & \makecell[c]{Both\\ST $\cap$ SDF} & \makecell[c]{Only\\ST} & \makecell[c]{Only\\SDF} \\
    \midrule
    \multirow{2}{*}{libxml2}
      & True Vulns.  & 38 & 10 & 10 &  8 & 13 &  7 & 11 \\
      & False Alarms & 38 & 11 & 10 &  9 & 11 & 10 & 13 \\
    \midrule
    \multirow{2}{*}{zstd}
      & True Vulns.  & 49 &  0 & 29 &  0 & 20 &  9 & 13 \\
      & False Alarms & 49 &  0 & 26 &  0 & 12 & 14 & 16 \\
    \midrule
    \multirow{2}{*}{PROJ}
      & True Vulns.  & 10 &  0 &  4 &  1 &  0 &  4 &  1 \\
      & False Alarms & 10 &  0 &  7 &  1 &  0 &  7 &  2 \\
    \midrule
    \textbf{Total} & -- & \textbf{194} & \textbf{21} & \textbf{86} & \textbf{19} & \textbf{56} & \textbf{51} & \textbf{56} \\
    \bottomrule
  \end{tabular}
  \label{table:intersection_comparison}
\end{table}


\subsection{RQ6: \RQsix}
\label{res6}

\noindent
\textit{\underline{Goal and Approach}}. 
This RQ compares two different forms of warning triage, \fs and LLM4SA~\cite{LLM4SA}. \fs represents the execution-based setting: it generates sliced binaries, fuzzes them, and uses sanitizer feedback to observe whether a warning is triggered, reached without failure, or not reached. LLM4SA represents the prediction-based setting: it reasons statically over code context from learned buggy patterns and predicts whether a warning is likely true or false without executing the program. Therefore, this comparison should not be interpreted as a direct tool-equivalence comparison, but as a study of the trade-offs between runtime evidence and static prediction for warning classification.

For \fs we enable all optimizations and compare results on dataset-TP and dataset-FP. We share the project wise classification results in Table~\ref{tab:llm_comparison}. The results demonstrate that \fs outperforms LLM4SA across all accuracy metrics for each project. Furthermore, we show the confusion matrices for \fs and LLM4SA in Table~\ref{tab:llm_confusion}. The confusion matrix indicates that LLM4SA is biased toward classifying warnings as false alarms (predicted 51 of cases in dataset-TP as false alarm and 47 of cases in dataset-FP as false alarm), which explains its comparatively better performance on the dataset-FP than on the dataset-TP. Overall, \fs outperforms LLM4SA, with an F1 score of 0.791 versus 0.360.

Moreover, from Table~\ref{tab:llm_comparison} we found that LLM4SA does not exhibit a substantial distinguishing power between the same code locations in dataset-TP and dataset-FP. This behavior arises from the construction of the datasets using the RevBugBench framework~\cite{Zhang2022}, in which different guard conditions are manipulated: stronger guards yield the dataset-FP (false alarms), whereas weaker guards result in the dataset-TP (true bugs). LLM4SA reasons about static analysis warnings by extracting a small code snippet from the vulnerable function. However, the guard that determines whether the vulnerability is exercised may reside in a transitive caller, preventing the LLM from reliably distinguishing true bugs from false alarms. In contrast, \fs incrementally expands code slices (up to three levels in our current work) based on reachability and vulnerability outcomes, thereby incorporating relevant guard conditions into the analysis.

These results should be interpreted in light of the differences in task formulation. \fs benefits from concrete runtime evidence produced by fuzzing and sanitizers, while LLM4SA is limited to the static context provided to the model. Conversely, LLM4SA does not require slice construction, compilation, or fuzzing infrastructure. Thus, our results do not imply that execution-based triage universally replaces prediction-based approaches. Rather, they show that, in our benchmark, sliced execution provides stronger evidence for distinguishing true vulnerabilities from false alarms, particularly when the decisive guard conditions are located outside the local code context available to the LLM.


\begin{table}[t]
\centering
\caption{Comparison of \fs and LLM4SA across projects.}
\label{tab:llm_comparison}
\begin{tabular}{llc cc cc}
\toprule
\multirow{2}{*}{Project} & \multirow{2}{*}{Dataset} & \multirow{2}{*}{Total} &
\multicolumn{2}{c}{\textbf{SnipTest}} &
\multicolumn{2}{c}{\textbf{LLM4SA}} \\
\cmidrule(lr){4-5} \cmidrule(lr){6-7}
 & & & TP & FP & TP & FP \\
\midrule

\multirow{2}{*}{libxml2}
 & True Vulns. & 38 & 20 & 0 & 12 & 15 \\
 & False Alarms & 38 & 7 & 21 & 14 & 14 \\
\midrule

\multirow{2}{*}{zstd}
 & True Vulns. & 49 & 29 & 0 & 6 & 32 \\
 & False Alarms & 49 & 20 & 26 & 8 & 30 \\
 \midrule

 \multirow{2}{*}{PROJ}
 & True Vulns. & 10 & 4 & 0 & 4 & 4 \\
 & False Alarms & 10 & 1 & 7 & 5 & 3 \\

\bottomrule
\end{tabular}
\end{table}

\begin{table}[tb!]
\centering
\caption{Confusion matrices for \fs and LLM4SA}
\label{tab:llm_confusion}
\resizebox{\columnwidth}{!}{%
\begin{tabular}{c cc c cc}
\toprule
& \multicolumn{2}{c}{\textbf{SnipTest}} & &
\multicolumn{2}{c}{\textbf{LLM4SA}} \\
\cmidrule(lr){2-3} \cmidrule(lr){5-6}

& \textbf{Pred True} & \textbf{Pred False} & &
\textbf{Pred True} & \textbf{Pred False} \\
\midrule

\textbf{Real True}
& 53 & 0 & & 22 & 51 \\

\textbf{Real False}
& 28 & 54 & & 27 & 47 \\

\bottomrule
\end{tabular}}
\end{table}

\subsection{RQ7: \RQseven{}}
\label{res7}

\noindent
\noindent
\textit{\underline{Goal and Approach}}. \textcolor{black}{This RQ compares \fs with the slice-based approach \oldfs. \fs builds upon \oldfs as its low-level slice-compilation component, but the two systems operate at fundamentally different units of reasoning: \oldfs analyzes a single Level‑0 slice, whereas \fs evaluates a family of caller-rooted slices organized by call-graph distance. In \fs, slices are fuzzed independently, observations are aggregated within each level, and the sequence of level outcomes is then synthesized into warning-level evidence. We compare \oldfs's single-slice output against \fs's full multi-level framework, reflecting their respective units of analysis.}


\textcolor{black}{We present the project-wise classification results in Table~\ref{tab:sniptest_vs_fuzzslice_slices}. As shown in the table, \oldfs is evaluated only at the Level 0 setting, since it cannot be directly configured to generate higher-level slices.
Both approaches exhibit comparable performance on PROJ. 
However, on projects such as libxml2 and zstd, \fs achieves higher precision across both datasets.
On the dataset of reached bugs, \fs achieves a precision, recall, and F1 score of 0.654, 1.0, and 0.791, respectively, by leveraging all three slice levels for warning classification. 
In comparison, \oldfs, using only the Level 0 slice for classification, achieves corresponding scores of 0.566, 1.0, and 0.722 on its reached bugs.
Both techniques have high recall as true vulnerabilities in dataset-TP are easily exploited by the fuzzer at the Level 0 slice when reached.
\fs achieves improvement in F1 score over \oldfs, primarily due to the inclusion of additional code context across slice levels, which enables more accurate classification of benign locations.}

\textcolor{black}{Finally, at Level 0 alone, \fs reaches and classifies 169 out of 194 locations, whereas \oldfs reaches and classifies 137. 
This corresponds to \fs reaching and classifying 32 additional locations, or 23.3\% more locations than \oldfs at Level 0 indicating improved wrapper generation for \fs as described in Approach Section~\ref{sec:approach}.}

\begin{table}[tb]
  \scriptsize
  \caption{
    Comparison of \fs and \oldfs across slice levels. 
    \textbf{V} = Bug oracle is triggered at the location in at least one slice at the level, 
    \textbf{B} = Location is reached in at least one slice but the bug oracle is not alerted,
    \textbf{NR} = Location is not reached in any slice at the level.
  }
  \centering
  \setlength{\tabcolsep}{3pt}
  \renewcommand{\arraystretch}{1.1}

  \begin{tabular}{llc r *{6}{r}}
    \toprule
    \multirow{2}{*}{Project} &
    \multirow{2}{*}{Dataset} &
    \multirow{2}{*}{\# Cases} &
    \multirow{2}{*}{Slice} &
    \multicolumn{3}{c}{\textbf{SnipTest}} &
    \multicolumn{3}{c}{\textbf{FuzzSlice}} \\
    
    \cmidrule(lr){5-7}
    \cmidrule(lr){8-10}

    & & & & V & B & NR & V & B & NR \\
    \midrule

    \multirow{6}{*}{libxml2}
      & \multirow{3}{*}{True Vulns.}
      & \multirow{3}{*}{38}
      & 0 & 30 & 0 & 8  & 20 & 0  & 18 \\
      & & & 1 & 26 & 0 & 12 & -  & -  & -  \\
      & & & 2 & 20 & 0 & 18 & -  & -  & -  \\

      \cmidrule{2-10}

      & \multirow{3}{*}{False Alarms}
      & \multirow{3}{*}{38}
      & 0 & 14 & 17 & 7  & 13 & 16 & 9 \\
      & & & 1 & 11 & 14 & 13 & - & - & - \\
      & & & 2 & 7  & 12 & 19 & - & - & - \\

    \midrule

    \multirow{6}{*}{zstd}
      & \multirow{3}{*}{True Vulns.}
      & \multirow{3}{*}{49}
      & 0 & 47 & 0 & 2  & 34 & 0  & 15 \\
      & & & 1 & 38 & 0 & 11 & -  & -  & - \\
      & & & 2 & 29 & 0 & 20 & -  & -  & - \\

      \cmidrule{2-10}

      & \multirow{3}{*}{False Alarms}
      & \multirow{3}{*}{49}
      & 0 & 30 & 13 & 6  & 30 & 9 & 10 \\
      & & & 1 & 24 & 18 & 7  & - & - & - \\
      & & & 2 & 20 & 26 & 3  & - & - & - \\

    \midrule

    \multirow{6}{*}{PROJ}
      & \multirow{3}{*}{True Vulns.}
      & \multirow{3}{*}{10}
      & 0 & 9 & 0 & 1 & 6 & 0 & 4 \\
      & & & 1 & 7 & 0 & 3 & - & - & - \\
      & & & 2 & 4 & 0 & 6 & - & - & - \\

      \cmidrule{2-10}

      & \multirow{3}{*}{False Alarms}
      & \multirow{3}{*}{10}
      & 0 & 3 & 6 & 1 & 3 & 6 & 1 \\
      & & & 1 & 1 & 5 & 4 & - & - & - \\
      & & & 2 & 1 & 0 & 9 & - & - & - \\

    \bottomrule
  \end{tabular}

  \label{tab:sniptest_vs_fuzzslice_slices}
\end{table}

\section{\fs on open source projects}
\label{sec:opensource}

\noindent
In the previous section, we evaluated \fs on a synthetic dataset from RevBugbench framework~\cite{Zhang2022}. 
In this section, we demonstrate the practical application of \fs by deploying our cloud-based \fs setup on two real-world open-source projects: Vim~\cite{vim-github} and libpcap~\cite{libpcap-github}. 
Vim is a widely used text editor, and libpcap is a common Linux utility for network packet capture. 

Using the static analysis tool Flawfinder~\cite{flawfinder}, we generated 2,328 warnings for Vim and 1,284 for libpcap.
From these, we selected warnings corresponding to high-impact vulnerabilities, such as CWE-120 (buffer overflow) and CWE-476 (null pointer dereference), 
which were not covered by native tests.
This process yielded 24 warnings in Vim and 9 in libpcap, which we then attempted to validate using \fs.

We fuzzed the associated multi-level code slices (levels 0--2) containing the sampled warnings. 
Out of the 33 candidate warnings, only four produced consistent crashes across all slice levels. 
For these, we submitted issues and pull requests to the affected projects. Within a week, this effort led to the confirmation of three new vulnerabilities across both projects: two buffer overruns and one null pointer dereference~\cite{vim-buffer,vim-null,libpcap-bug}. The fourth case, an out-of-bounds write in Vim, was disputed by developers, who argued in an email thread that the behavior was intentional and not a bug. We discuss each identified vulnerability and its fix below:
\begin{enumerate}
    \item \textbf{Buffer overflow in libpcap~\cite{libpcap-bug}.} Libpcap is a portable API for low-level network monitoring and traffic analysis across Unix-like systems and Windows (via WinPcap). A heap buffer overflow was identified in the function utf\_16le\_to\_utf\_8\_truncated, which is responsible for converting Windows error messages from UTF-16 to UTF-8. During this conversion, the function incorrectly counts the number of characters written when determining how much data to truncate, resulting in a heap buffer overwritten at the end of the destination buffer. This vulnerability was assigned CVE-2025-11964~\cite{CVE}. The developer’s fix corrects the character counting logic used during conversion. The developer released a new version of the software patching the vulnerability and appreciating the authors for spotting the issue in the release notes~\cite{CVE}.
    \item \textbf{Buffer overflow in Vim~\cite{vim-buffer}.} Vim is a highly configurable text editor designed for efficient text creation and modification. A buffer overflow was discovered in Vim within the function cs\_pathcomponents, which truncates file paths based on a user-supplied integer value (e.g., displaying only the last N directories of a path). The truncation logic contains an error in string pointer manipulation—specifically, a double decrement—leading to a buffer underflow. The issue was fixed by correcting the pointer logic to perform a single decrement instead of two.
    \item \textbf{Null dereference in Vim~\cite{vim-null}.} A null pointer dereference was identified in Vim in the function cs\_find\_common, which is executed when the find command is invoked within cscope. During execution, the function attempts to create a temporary file. If this operation fails (e.g., due to insufficient disk space, filename conflicts, or inadequate permissions), the resulting null pointer may still be accessed later in the function, causing a crash. The fix applied for the bug ensures that the function exits early if temporary file creation fails, preventing dereference of a null pointer.
\end{enumerate}

The remaining warnings either did not produce crashes at all slice levels or were not reachable (categorized as PFP or NR). 
While our analysis does not constitute a full-scale evaluation of several static analysis warnings, we intended to demonstrate the practical utility of \fs in identifying real vulnerabilities in widely-used open-source software.

\section{Limitations and Future Work}

We now discuss an exhaustive list of factors in the \fs framework that can create false positives and false negatives. We also discuss potential mitigation strategies.

\begin{enumerate}
    \item \textbf{Incomplete minimal context for bugs (false positives only)}. A reported bug may be classified as a true vulnerability or a false alarm based on a guard condition that resides several functions earlier in the call hierarchy. In our current implementation, \fs limits analysis to at most three slice levels. Extending the analysis to additional slice levels could incorporate such transitive guard conditions and help avoid misclassifying false alarms as true vulnerabilities.
    \item \textbf{Insufficient fuzzing time relative to slice size (false negatives or unreached code)}. In some cases, the allocated fuzzing time is insufficient to adequately explore larger code slices. In our evaluation, the time budget is 5/7.5/10 minutes for Levels 0, 1 and 2. These limits are empirically chosen and may be inadequate for certain complex slices. Moreover, large test cases (e.g., functions with many or large arguments) and inefficient mutation strategies can further hinder exploration. Such limitations may result in unreached code or false negatives - where reachable bugs are not exploited within the allotted time. For example, Table~\ref{tab:repetitions} shows 5 repetitions of \fs on our dataset. We observed 2 cases in libxml2 (dataset-TP) crashing at all slice levels on some iterations and not being reached in others. However in dataset-FP, the same cases can be reached  at higher slice levels, classifying them as benign. 
    \item \textbf{Function pointers, file I/O, and runtime environment variables (false positives, false negatives or unreached code).} \fs currently assigns function pointers to NULL and does not randomly bind them to functions within the slice. In addition, it does not model file-based I/O or runtime environment variables that influence program behavior. These limitations can lead to both false positives, false negatives or unreached code. Addressing these aspects is left for future work.
\end{enumerate}

\noindent
Finally, we note that the cross-level classification rule in \fs is heuristic. Although it is empirically motivated by the observed stability of warning behavior across slice levels, it does not provide formal soundness or completeness guarantees. A warning may still be misclassified if the generated slices omit relevant program context, if required environmental state is not modeled, or if the fuzzer fails to explore the necessary inputs within the time budget. We therefore interpret output of \fs  as evidence for prioritization rather than definitive validation.

\begin{table}[tb!]
\centering
\caption{\fs results across projects over repeated runs}
\label{tab:repetitions}
\begin{tabular}{lccc}
\toprule
\# & All TP & All PFP & All NR \\
\midrule

\#1 (True Vulns.) & 53 & 0  & 44 \\
\#2 (True Vulns.) & 53 & 0  & 44 \\
\#3 (True Vulns.) & 51 & 0  & 46 \\
\#4 (True Vulns.) & 53 & 0  & 44 \\
\#5 (True Vulns.) & 52 & 0  & 45 \\
\midrule
\#1 (False alarms) & 28 & 54 & 15 \\
\#2 (False alarms) & 28 & 54 & 15 \\
\#3 (False alarms) & 28 & 54 & 15 \\
\#4 (False alarms) & 28 & 54 & 15 \\
\#5 (False alarms) & 28 & 54 & 15 \\
\bottomrule
\end{tabular}
\end{table}

\section{Conclusion}
\textcolor{black}{We presented \fs, an execution-based warning-triage framework built atop the \oldfs function-slice compilation primitive. \fs organizes slices into progressively broader context levels and aggregates warning behavior both within and across these levels. This methodology yields evidence about whether a warning persists as caller-imposed constraints are gradually restored.
In our benchmark, \fs produced Possible True Positive evidence for 53 of 97 confirmed vulnerabilities and Possible False Positive evidence for 54 of 97 confirmed false alarms when analyzing across three slice levels. It also generated Possible True Positive evidence for 28 false alarms, underscoring that insufficient context remains an important limitation. \fs should therefore be viewed as a prioritization aid that complements developer review.
Our evaluation further demonstrates that sliced execution can substantially reduce fuzzing time and expose previously unknown defects, as evidenced by our disclosure of CVE-2025-11964.
These results support multi-level sliced fuzzing as a practical approach for gathering warning-triage evidence—rather than as definitive proof of vulnerability or benignness.}


 \def\UrlBreaks{\do\/\do-}
\balance
\bibliographystyle{IEEEtranS}
\balance
\bibliography{Bibliography}

\vskip -2\baselineskip plus -1 fil
\begin{IEEEbiography}
[{\includegraphics[width=1.5in,height=1in,clip,keepaspectratio]{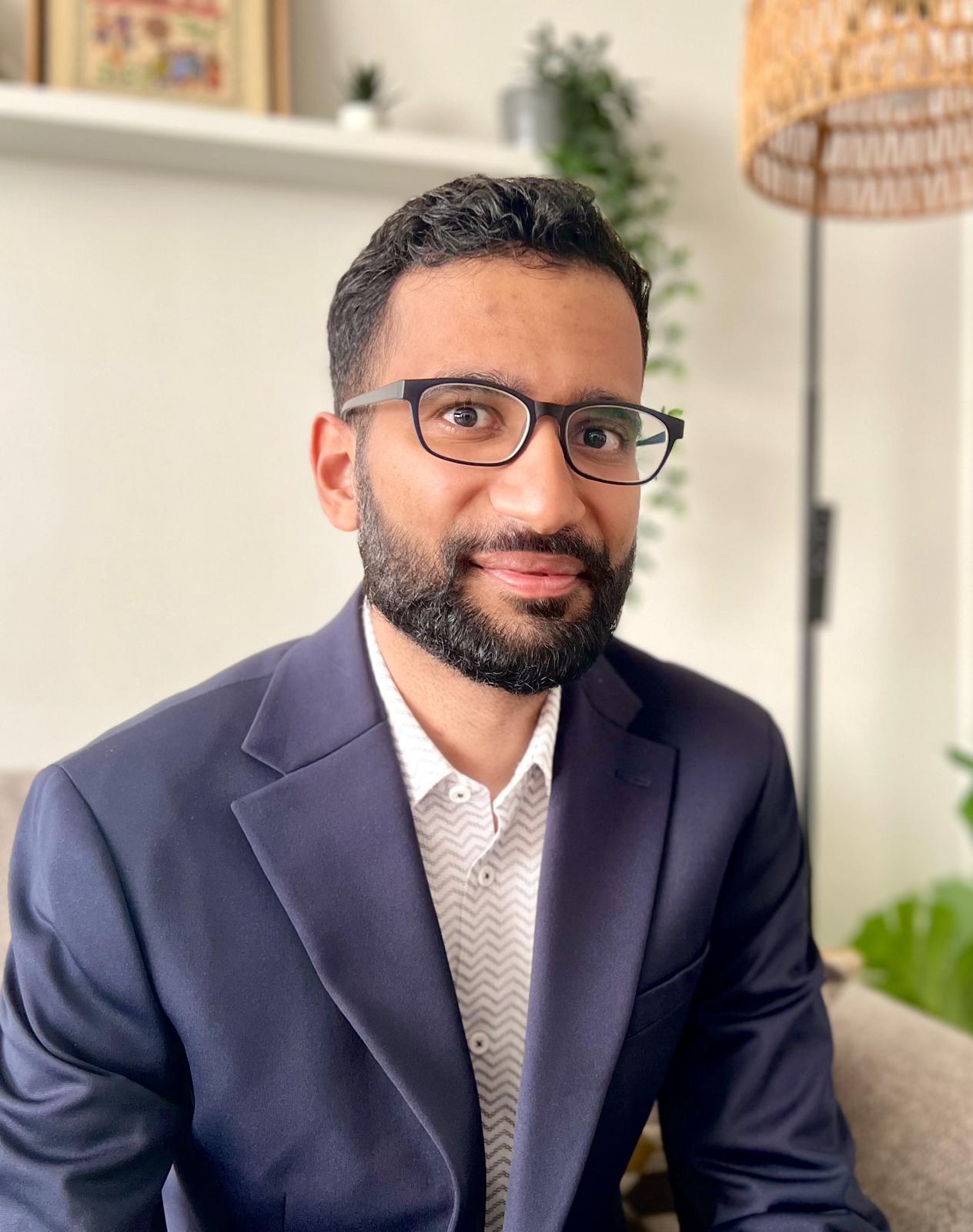}}]
{Aniruddhan Murali} is a Ph.D. graduate from the
Cheriton School of Computer Science at the
University of Waterloo, Canada. His research
interests include fuzzing, code slicing, vulnerability detection, and automatic bug fixing. You can find more about him \href{https://www.linkedin.com/in/aniruddhan-murali-40b41a11a/}{here}.
\end{IEEEbiography}

\vskip -2\baselineskip plus -1 fil
\begin{IEEEbiography}
[{\includegraphics[width=1.5in,height=1in,clip,keepaspectratio]{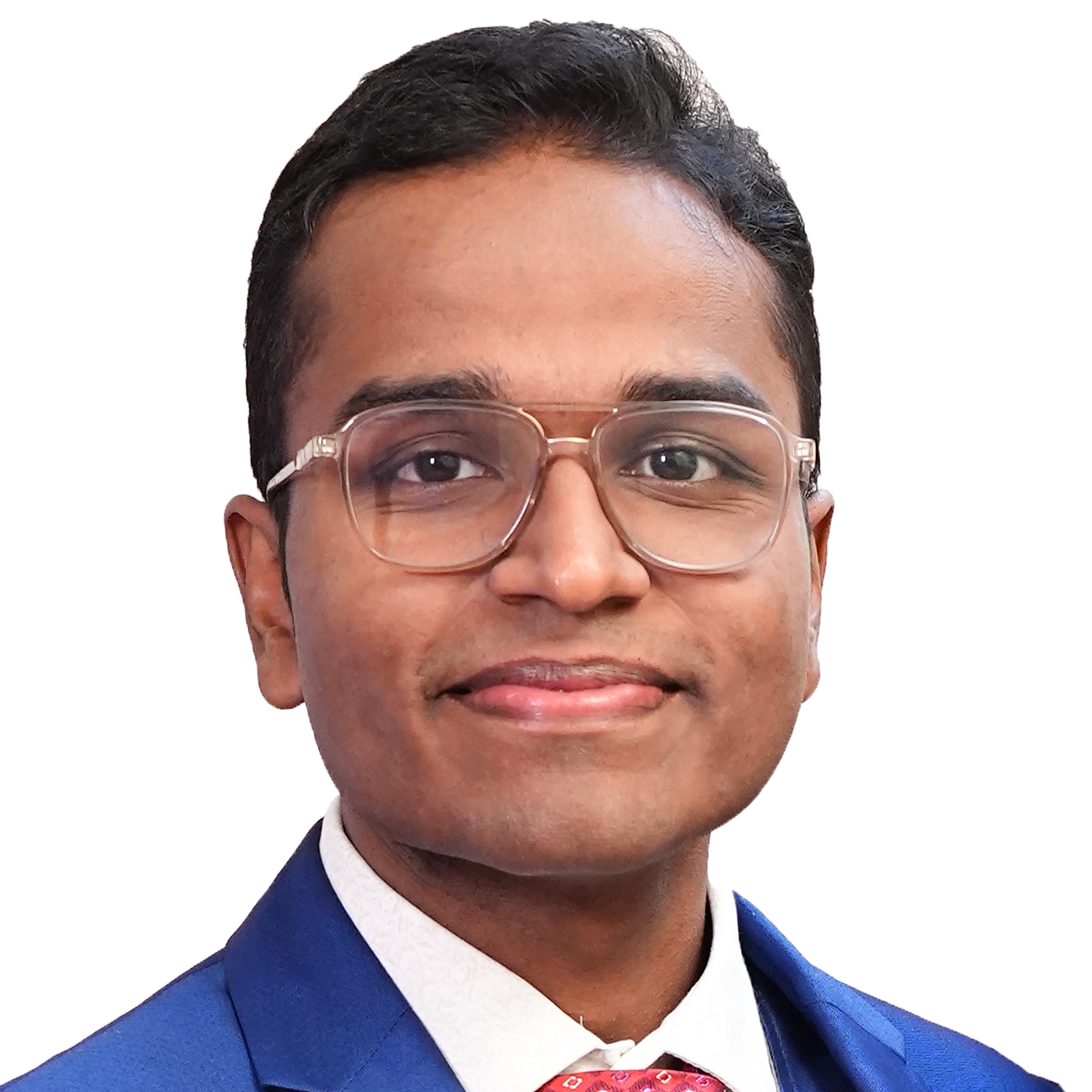}}]
{Noble Saji Mathews} is a Research Associate in the David R. Cheriton School of Computer Science at the University of Waterloo. His research interests include artificial intelligence and software engineering. Noble received his Masters in Computer Science from the University of Waterloo. You can find more about him \href{https://noblesm.com}{here}.
\end{IEEEbiography}

\vskip -4\baselineskip plus -1 fil
\begin{IEEEbiography}
[{\includegraphics[width=1.5in,height=1in,clip,keepaspectratio]{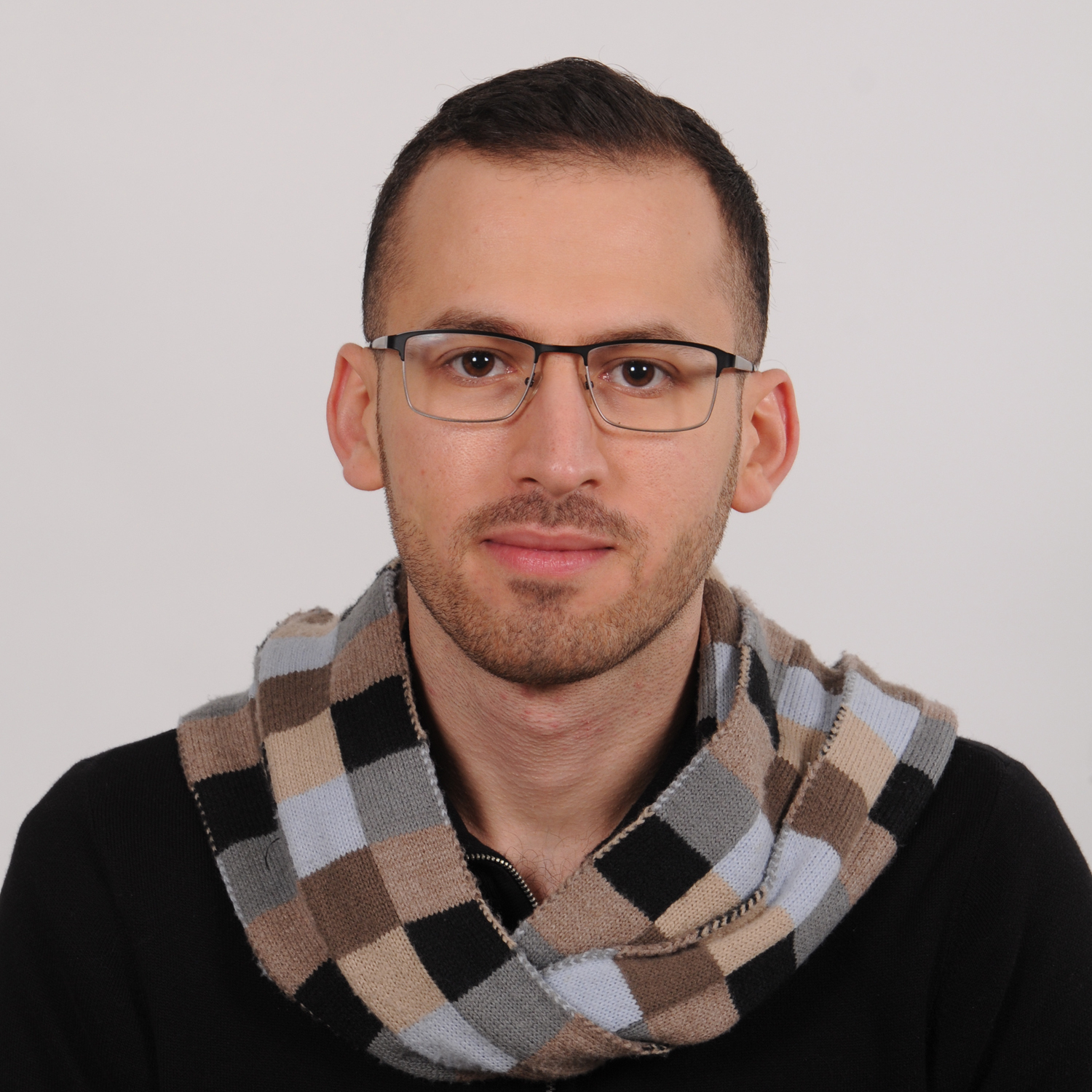}}]{Mahmoud Alfadel} is an Assistant Professor with the Department of Computer Science, University of Calgary. His research interests include software ecosystems, software security, and release engineering.
\end{IEEEbiography}

\vskip -4\baselineskip plus -1fil
\begin{IEEEbiography}
[{\includegraphics[width=1.5in,height=1in,clip,keepaspectratio]{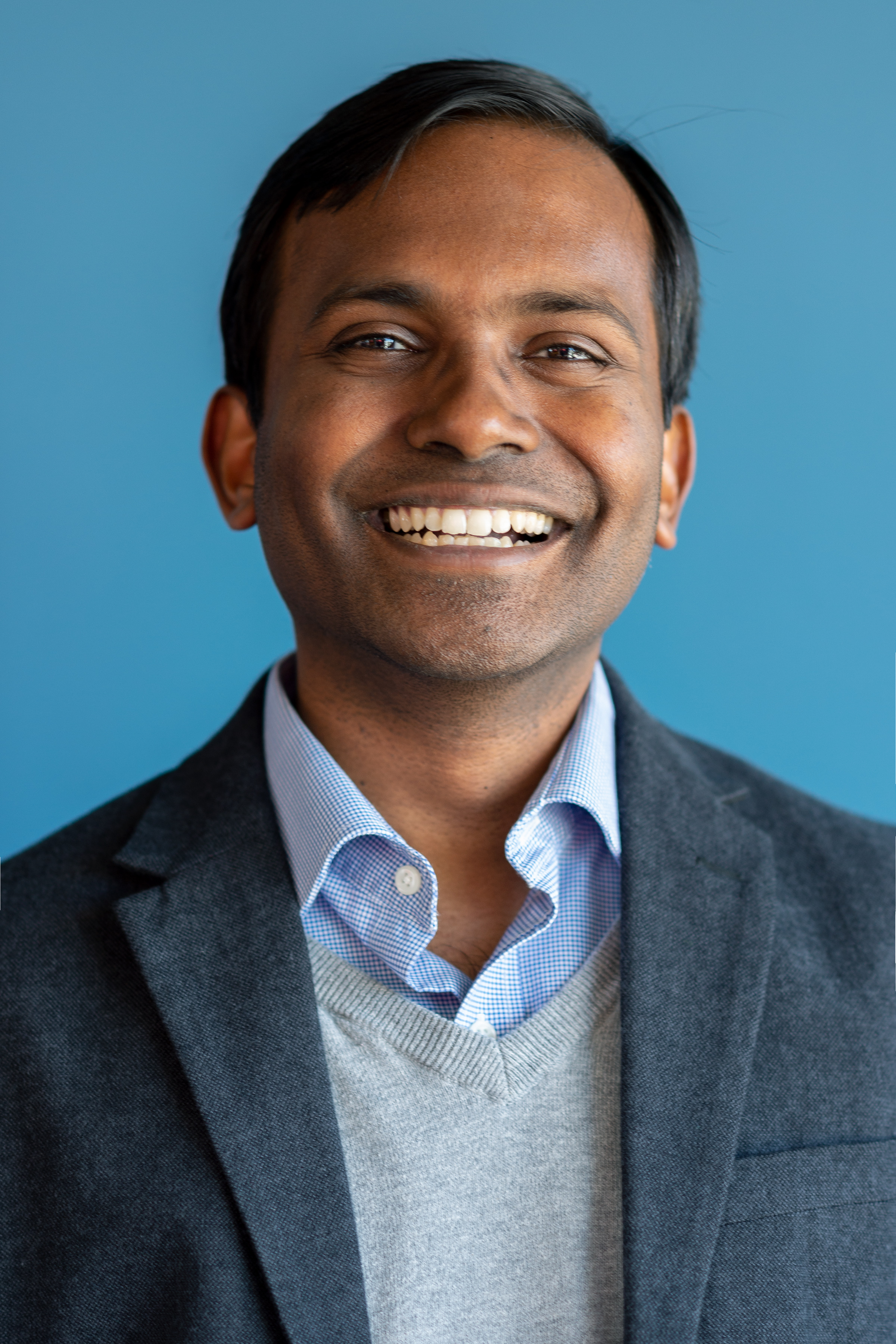}}]{Meiyappan Nagappan} is an Associate Professor at the Cheriton School of Computer Science, University of Waterloo. He has worked on empirical software engineering to address software development concerns and currently researches the impact of large language models on software development.
\end{IEEEbiography}

\vskip -4\baselineskip plus -1fil
\begin{IEEEbiography}
[{\includegraphics[width=1.5in,height=1in,clip,keepaspectratio]{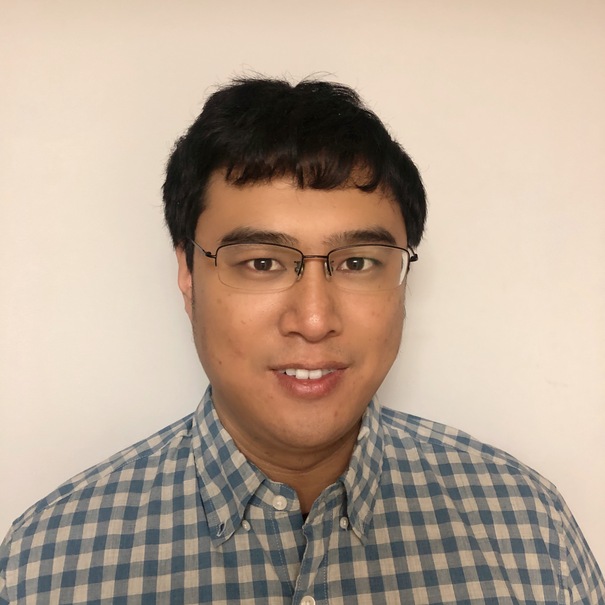}}]{Meng Xu} is an Assistant Professor in the Cheriton School of
Computer Science at the University of Waterloo, Canada. His research is
in the area of system and software security, with a focus on delivering
high-quality solutions to practical security programs, especially in
finding and patching vulnerabilities in critical computer systems. This
usually includes research and development of automated program analysis
/ testing / verification tools that facilitate the security reasoning
of critical programs.
\end{IEEEbiography}




\end{document}